\documentclass[
preprint,
 amssymb, amsmath,
 aps,cha,
titlepage,
longbibliography
]{revtex4-1}

\usepackage{graphicx}
\usepackage{epstopdf}
\usepackage{mathtools}
\usepackage[percent]{overpic}
\usepackage[colorlinks=true,linkcolor=blue]{hyperref}
\usepackage{subcaption}
\DeclareCaptionFormat{cont}{#1 (cont.)#2#3\par}

\draft

\begin{document}

\title{Shock waves from non-spherical cavitation bubbles}

\author{Outi Supponen}
\email[]{outi.supponen@epfl.ch}
\affiliation{Laboratory for Hydraulic Machines, Ecole Polyt\'echnique F\'ed\'erale de Lausanne, Avenue de Cour 33bis, 1007 Lausanne, Switzerland}
\author{Danail Obreschkow}
\affiliation{International Centre for Radio Astronomy Research, University of Western Australia, 7 Fairway, Crawley, WA 6009, Australia}
\author{Philippe Kobel}
\affiliation{Laboratory for Hydraulic Machines, Ecole Polyt\'echnique F\'ed\'erale de Lausanne, Avenue de Cour 33bis, 1007 Lausanne, Switzerland}
\author{Marc Tinguely}
\affiliation{Laboratory for Hydraulic Machines, Ecole Polyt\'echnique F\'ed\'erale de Lausanne, Avenue de Cour 33bis, 1007 Lausanne, Switzerland}
\author{Nicolas Dorsaz}
\affiliation{Laboratory for Hydraulic Machines, Ecole Polyt\'echnique F\'ed\'erale de Lausanne, Avenue de Cour 33bis, 1007 Lausanne, Switzerland}
\author{Mohamed Farhat}
\affiliation{Laboratory for Hydraulic Machines, Ecole Polyt\'echnique F\'ed\'erale de Lausanne, Avenue de Cour 33bis, 1007 Lausanne, Switzerland}

\date{\today}

\begin{abstract}
We present detailed observations of the shock waves emitted at the collapse of single cavitation bubbles using simultaneous time-resolved shadowgraphy and hydrophone pressure measurements.
The geometry of the bubbles is systematically varied from spherical to very non-spherical by decreasing their distance to a free or rigid surface or by modulating the gravity-induced pressure gradient aboard parabolic flights.
The non-spherical collapse produces multiple shocks that are clearly associated with different processes, such as the jet impact and the individual collapses of the distinct bubble segments.
For bubbles collapsing near a free surface, the energy and timing of each shock are measured separately as a function of the anisotropy parameter $\zeta$, which represents the dimensionless equivalent of the Kelvin impulse.
For a given source of bubble deformation (free surface, rigid surface or gravity), the normalized shock energy depends only on $\zeta$, irrespective of the bubble radius $R_{0}$ and driving pressure $\Delta p$. 
Based on this finding, we develop a predictive framework for the peak pressure and energy of shock waves from non-spherical bubble collapses.
Combining statistical analysis of the experimental data with theoretical derivations, we find that the shock peak pressures can be estimated as jet impact-induced hammer pressures, expressed as $p_{h} = 0.45\left(\rho c^{2}\Delta p\right)^{1/2} \zeta^{-1}$ at $\zeta > 10^{-3}$.
The same approach is found to explain the shock energy quenching as a function of $\zeta^{-2/3}$.
\end{abstract}

\maketitle 

\section{Introduction}


Shock waves are one of the most destructive phenomena occurring during the collapse of cavitation bubbles, and therefore a topic of long-standing interest.
The associated pressures, reaching values in the order of GPa~\cite{Pecha2000,Akhatov2001}, are able to wear metallic surfaces, which is a classic concern for ship propellers and hydraulic turbines~\cite{Silverrad1912,Arndt1981,Vogel1989,Escaler2006}.
Further victims of cavitation-induced damage are, for example, artificial heart valves~\cite{Rambod1999}, liquid-propelled rocket engines~\cite{Jakobsen1971} and the prey of a mantis shrimp~\cite{Patek2005}.
The damaging power can also be exploited for beneficial uses such as in medical~\cite{Brennen2015} (e.g. shock wave lithotripsy~\cite{Field1991,Sass1991}, cancer therapy~\cite{Yu2004,Brennen2015}) and cleaning~\cite{Song2004} applications.
However, predictive tools to characterize the key properties of cavitation-driven shocks are limited.
In the quest of mitigating the harm they may cause or maximizing their benefit, we here make detailed observations of shocks of single cavitation bubbles and propose a framework to predict their `strengths'.

Much progress has been made in the prediction of the damage potential of shock waves emitted by spherically collapsing bubbles~\cite{Hickling1964,Fujikawa1979,Akhatov2001,Fuster2010,Magaletti2015}.
However, doing so for non-spherically collapsing bubbles is still an open problem.
Bubbles may deform under the effect of, for example, nearby surfaces, inertial forces such as gravity or passing shock waves.
The collapse shock wave strengths have been shown, both experimentally and numerically, to vary with the bubble sphericity for bubbles collapsing near a rigid wall~\cite{Vogel1988,Ohl1999,Hsiao2014,Wang2016b}.
Shocks from bubbles collapsing under the effect of a passing shock wave have been shown sensitive to the latter's timing and strength~\cite{Sankin2005}.
The shocks emitted at the collapse of an individual bubble are often referred to as a single event, yet it is known that deformed bubbles that are pierced by high-speed micro-jets produce several shock waves from multiple locations upon collapse~\cite{Ohl1999,Lindau2003,Supponen2015}.
However, understanding the contribution of each shock emission mechanism to the final damage characteristics and a systematic study on the influence of the bubble deformation on them is still lacking, as recently was pointed out by Lauterborn~\cite{Lauterborn2013}.
Although numerical simulations offer an excellent means to reproduce complex shock wave scenarios associated to non-spherical collapses~\cite{Johnsen2009,Chahine2015,Koukouvinis2016,Koukouvinis2016b}, observations for their validation are limited.
Furthermore, we still lack an understanding on how the shocks from bubbles deformed by distinct sources differ.

In this work, shock wave energies and pressures are systematically measured as a function of the various bubble parameters and asymmetries.
The objective is to understand how the deformation of bubbles affects their detailed collapse shock wave emission.
In particular, we aim to estimate, through visualizations and pressure measurements, the strengths and the timings of the distinct shock waves produced at the collapse of bubbles with geometries varying from highly spherical to strongly deformed by a nearby free surface.
These data are then compared to bubbles deformed by a nearby rigid surface and by the hydrostatic pressure gradient, which is modulated in variable gravity aboard parabolic flights (60th and 62nd European Space Agency parabolic flight campaigns and the 1st Swiss parabolic flight).
The advantage of a gravity-induced pressure gradient to deform bubbles is its uniformity in time and space that leads to similar bubble `collapse shapes' across a wide range of bubble asymmetries~\cite{Supponen2016}.
Furthermore, any smooth pressure field can be approximated to first order by such a uniform pressure gradient.
We exploit the large number of data and a broad parameter space to reach an empirical model for predicting the shock strengths for non-spherical bubbles, which is backed-up by theoretical arguments.
This model applies the scaling laws for micro-jets, which we have recently developed in detail~\cite{Supponen2016}, to the shock wave emission of deformed cavitation bubbles.

The deformation of bubbles collapsing near surfaces is usually quantified by the stand-off parameter $\gamma = h/R_{0}$, where $h$ is the distance between the bubble center and the surface and $R_{0}$ is the maximum bubble radius.
Deformations caused by near surfaces and gravity can be compared by using the vector-parameter $\boldsymbol{\zeta}$~\cite{Supponen2016,Obreschkow2011}:
\begin{equation}
\boldsymbol{\zeta} = \left \{
  \begin{array}{l l}
    -\rho\mathbf{g}R_{0}\Delta p^{-1} & \quad \text{gravitational field}\\
   +0.195\gamma^{-2}\mathbf{n}  & \quad \text{flat free surface}\\
   -0.195\gamma^{-2}\mathbf{n}  & \quad \text{flat rigid surface}
  \end{array} \right.
\label{eq:zeta}
\end{equation}
where $\rho$ is the liquid density, $\mathbf{g}$ is the gravitational acceleration, $\Delta p = p_{0}-p_{v}$ is the driving pressure (where $p_{0}$ is the static pressure of the unperturbed liquid at the location of the bubble and $p_{v}$ is the vapor pressure) and $\mathbf{n}$ is the unit vector normal to the surface, in the direction from the surface to the bubble. 
$\boldsymbol{\zeta}$ is essentially the dimensionless equivalent of the Kelvin impulse, which is the linear momentum acquired by the liquid during the growth and the collapse of the bubble~\cite{Blake1988}.
A higher $\zeta \equiv |\boldsymbol{\zeta}|$ causes a more pronounced bubble deformation and delineates key parameters of the micro-jet, such as the jet speed or the jet impact timing, almost irrespective of the source of deformation for $\zeta<0.1$~\cite{Supponen2016}.
We henceforth primarily use $\zeta$ to quantify bubble deformation, but also display the equivalent $\gamma$ for convenience.

This paper is structured as follows.
Firstly, Section~\ref{s:experiment} presents the experimental methods, describing the setup and the relevant calibrations.
Section~\ref{s:details} shows detailed observations of single and multiple shock waves emitted by bubbles near a free surface.
A framework for predicting shock peak pressures and energies is then proposed in Section~\ref{s:models}, along with comparisons between shocks from bubbles deformed by different sources (free/rigid surface and gravity).
Finally, the results are discussed in Section~\ref{s:discussion}.

\section{Experimental methods}
\label{s:experiment}

\begin{figure}[b]
\begin{center}
\includegraphics[width=\textwidth]{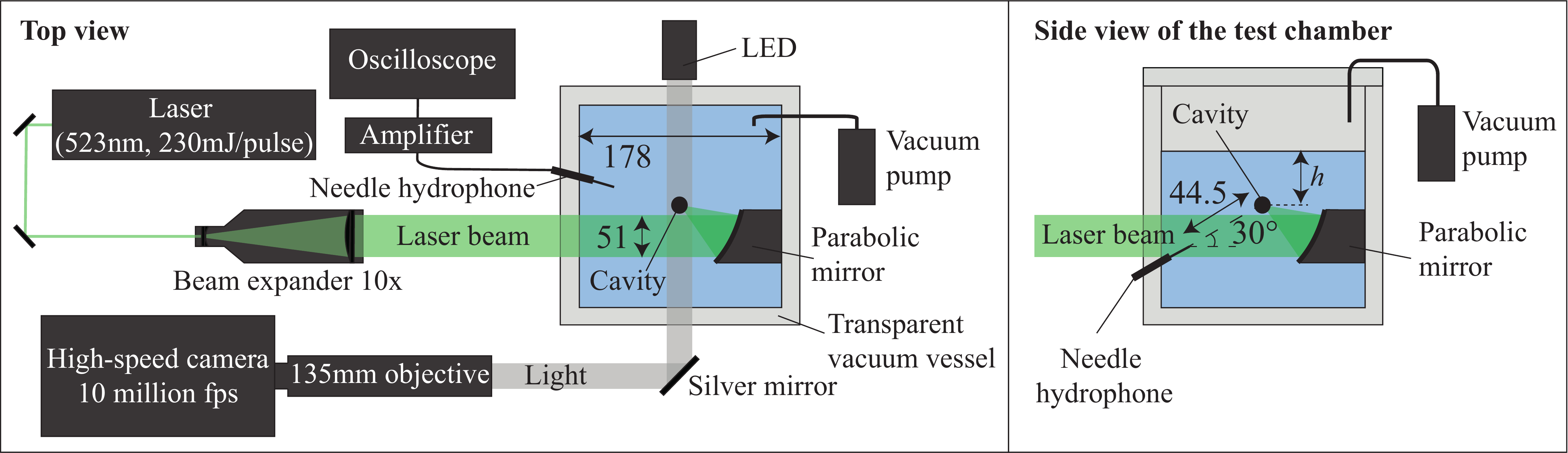}
\end{center}
\caption{Top and side view schematics of the experimental setup. The dimensions are given in mm.}
\label{fig:experiment}
\end{figure}
%
The central components of our experimental setup are shown in Fig.~\ref{fig:experiment}.
A pulsed laser is expanded and focused in demineralized water by an immersed parabolic mirror, which produces a point-like plasma and thereby an initially highly spherical bubble~\cite{Obreschkow2013} that grows and, subsequently, collapses.
The bubble and the associated shock waves are visualized using shadowgraphy with an ultra-high-speed camera (Shimadzu HPV-X2) reaching filming speeds up to 10~million frames per second with a 50~ns exposure time and a collimated backlight beam from a light emitting diode.
The driving pressure $\Delta p$ can be adjusted by varying the static pressure $p_{0}$ in the test chamber between 0.08 and 1~bar with a vacuum pump.
Tuning the laser power generates bubbles of energies $E_{0} = (4\pi/3)R_{0}^{3}\Delta p$ ranging from 0.1 to 28~mJ.
This parameter space leads to a wide range of maximum bubble radii, $R_{0}=1$--$10$~mm, which are large enough for viscosity and surface tension to have a negligible effect on the bubble dynamics~\cite{Levkovskii1968}.

To modulate the bubble deformation, we vary the bubble's distance to a surface ($h \sim 3$--$30$~mm) and/or the perceived gravity ($|\mathbf{g}|\sim 0$--$2$~$g$, where $g=9.81$~ms$^{-2}$), in addition to varying $R_{0}$ and $\Delta p$.
The maximum radii are obtained from the recorded collapse time $T_{c}$ (i.e.~half oscillation time) of the bubble as $R_{0}=1.093T_{c}(\Delta p/\rho)^{1/2}\kappa^{-1}$~\cite{Rayleigh1917}, where $\kappa$ is a factor depending on the source and level of deformation.
For bubbles collapsing near a free surface, $\kappa$ is a lifetime-shortening factor that can be approximated as $\kappa \approx 1-0.102\gamma^{-1}$~\cite{Gregorcic2007}.
The bubbles deformed by gravity or a nearby rigid surface in this work are at $\zeta<10^{-2}$, and therefore the deformations are weak enough for them to justify the assumption $\kappa\approx1$.
All measurements are made at room temperature.
Additional details on our experimental setup and the parabolic flights may be found in ref.~\citep{Obreschkow2013}.
%
\begin{figure}
\begin{center}
\includegraphics[width=.6\textwidth, trim=0.2cm 0cm 0.9cm 0.1cm, clip]{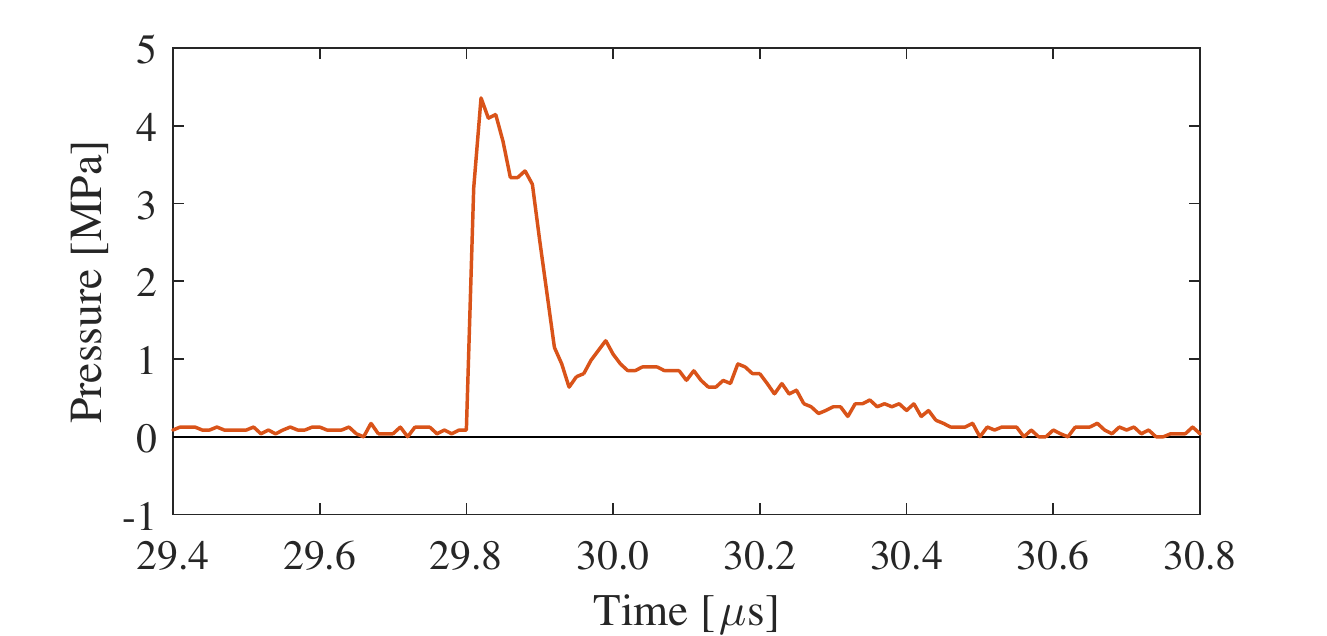}
\end{center}
\caption{A typical hydrophone pressure signal of the shock wave emitted at the bubble generation. $t=0$~$\mu$s corresponds to the time instant of bubble generation.}
\label{fig:gen}
\end{figure}

A needle hydrophone (75 $\mu$m sensor, manufactured by Precision Acoustics) is used to record the pressure of the shock waves.
The bandwidth of this hydrophone is guaranteed to extend above 30~MHz, and is thus capable of a detailed sampling of the shock waveform and of disentangling multiple fronts. 
The rise time upper bound is found to be approximately 15~ns, estimated from the time it takes for the pressure signal of the steep shock wave produced at the explosive bubble generation (Figure~\ref{fig:gen}) to rise from 10\% to 90\% of its maximum amplitude. 
The actual rise time of the shock wave is likely to be even shorter~\cite{Vogel1996}.
The pressure signal, represented by an electrical voltage, is amplified and recorded at 100~MHz sampling frequency by an oscilloscope.
The hydrophone sensor is located at a distance of $d = 44.5$~mm from the bubble center at an angle of 30$^{\circ}$ below the horizontal plane with a planar incidence of the shock wave onto the sensor.
The shock waves take approximately 30~$\mu$s to reach the hydrophone after being generated.
Being thin (needle thickness is 0.3~mm) and located far relative to the bubble size, the presence of the hydrophone needle is assumed to have a negligible effect on the bubble dynamics.
%
\begin{figure}
\begin{center}
\includegraphics[width=.9\textwidth]{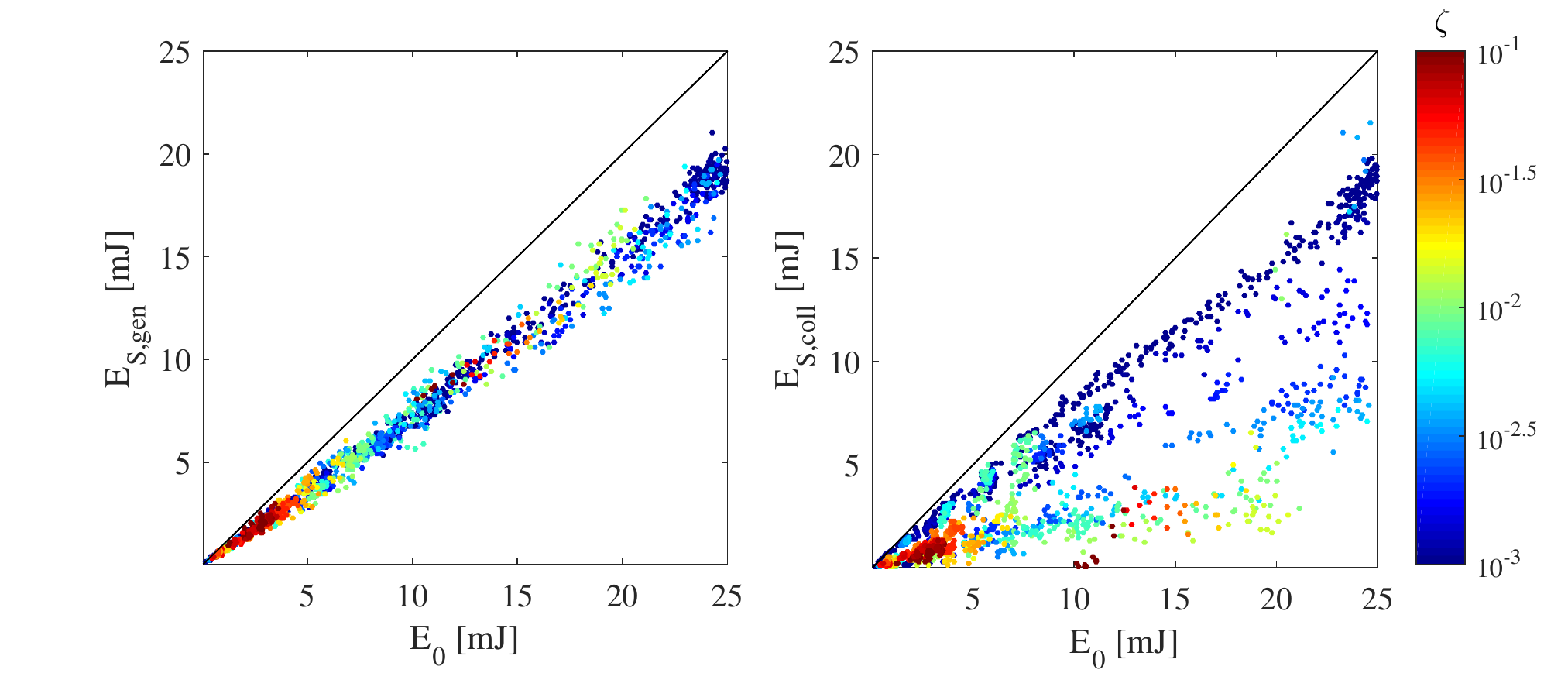}
\caption{Energies of shock waves emitted at bubble generation (left) and collapse (right) for various bubble energies $E_{0}$. The colors indicate the level of $\zeta$. The solid lines show $E_{S}=E_{0}$.}
\label{fig:E0_ES}
\end{center}
\end{figure}

We assume spherical propagation of the shock waves, and estimate their energies as
\begin{equation}
E_{S}=aU_{\rm max}^{b}\int U(t)^{2}{\rm d}t
\label{eq:ES}
\end{equation}
where $U(t)$~[V] is the hydrophone voltage signal (containing the full shock wave scenario in the case of multiple collapse shocks, but excluding any reflections from boundaries), $U_{\rm max}$ is the maximum value of $U(t)$ and $a$ and $b$ are calibration constants.
If the shock propagated with no energy dissipation, then $a=4\pi d^{2}\left(\rho c\right)^{-1}G^{-2}$~\cite{Vogel1988} (where $c$ is the sound speed in the liquid and $G$ is the gain in units of $[V/Pa]$) and $b=0$.
An exponent $b>0$ is used to approximately compensate for non-linear dissipation (e.g.~due to inelastic heating, induced micro-cavitation, etc.), whose relative effect increases with pressure.
As the precise gain $G$ is unknown in our current setup, and non-linear dissipation is expected, we treat $a$ and $b$ as positive free parameters.
We fit these parameters to simultaneously satisfy two conditions: (1) The energy of the laser-induced shock at the bubble generation $E_{\rm S,gen}$ scales linearly with the bubble energy $E_{0}$~\cite{Vogel1988}, and 
(2) The total energy of the shock(s) emitted at the bubble collapse $E_{\rm S,coll}$ is bounded by the difference between the bubble energy $E_{0}$ and the rebound energy $E_{\rm reb}$. 
For bubbles that collapse spherically ($\zeta < 10^{-3}$) and produce no jets, we assume $E_{\rm S,coll}\approx E_{0}-E_{\rm reb}$~\cite{Tinguely2012}.
We find that $a$ is such that $E_{\rm S,gen}/E_{0} \approx 0.75$ (i.e.~43\% of the absorbed laser energy goes into the generation shock and 57\% goes into the bubble) and $b\approx 0.45$, indicating slight non-linear dissipation.
Figure~\ref{fig:E0_ES} displays the calibrated energies both for bubble generation and collapse shocks waves for various $E_{0}$ and $\zeta$, clearly showing the linear relationship between $E_{\rm S,gen}$ and $E_{0}$ and that the collapse shock energies tend to be lower for increasing $\zeta$.
Pressures are then computed from the calibrated energies as $p(t)=U(t)/G$, where the gain $G$ is determined for each individual bubble separately as $G^{2}=4\pi d^{2}\left(\rho c\right)^{-1}\int U(t)^{2}{\rm d}t/E_{S}$.
Using a variable $G$ allows to compare the signals obtained in different conditions, for which the recorded pressures are differently affected by the shock's non-linear dissipation.

\section{Detailed observations}
\label{s:details}

\subsection{Spherical collapse}
\label{s:spherical}

\begin{figure}
\begin{center}
\begin{subfigure}{0.4\textwidth}
	\includegraphics[width=\textwidth]{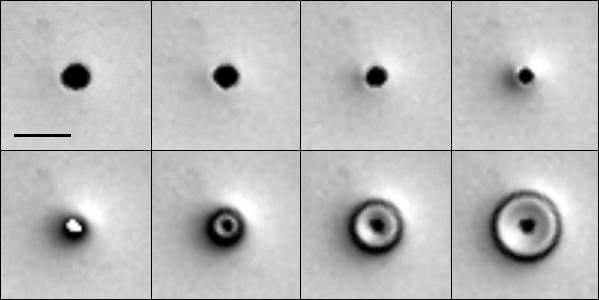}
	\caption{ }
	\label{fig:spherical_visu}
\end{subfigure}
\begin{subfigure}{0.58\textwidth}
	\includegraphics[width=\textwidth]{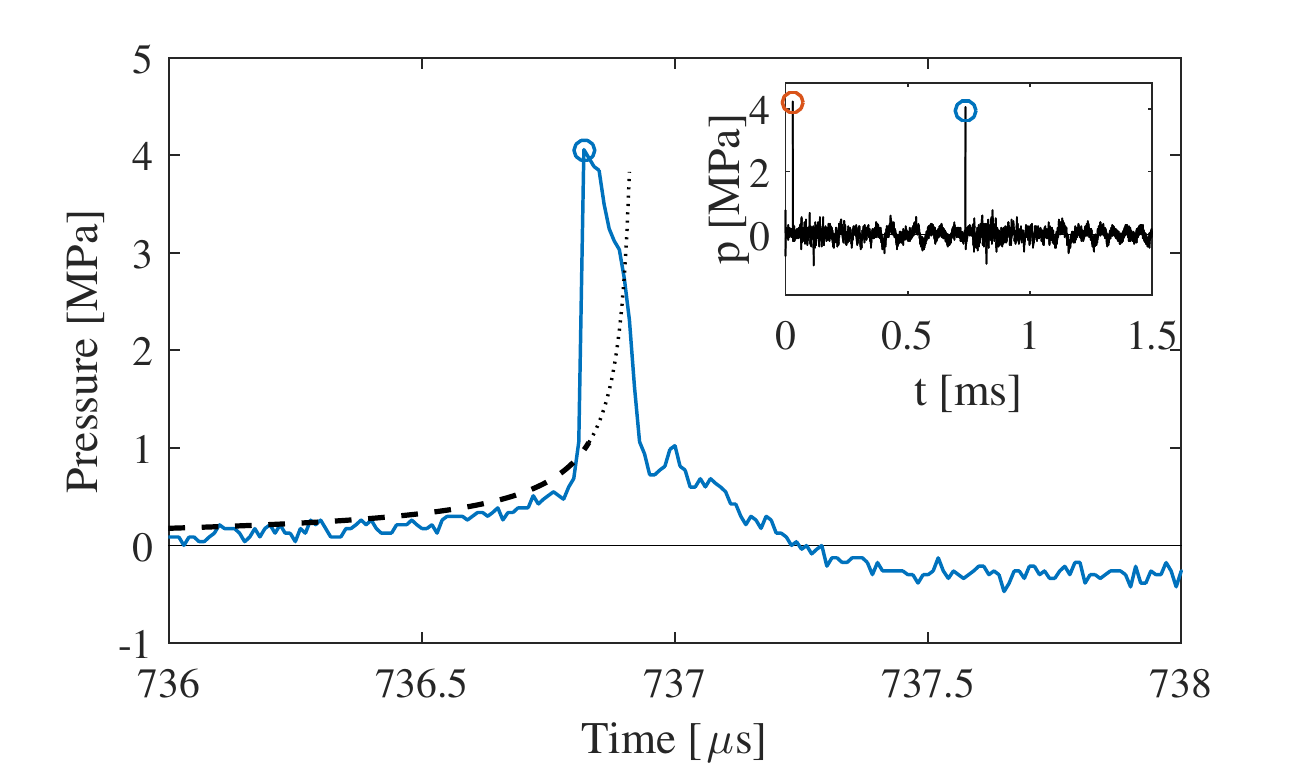}
	\caption{ }
	\label{fig:spherical_hydro}
\end{subfigure}
\end{center}
\caption{Cavity of $R_{0} = 3.8$~mm collapsing spherically at $\zeta < 10^{-3}$ and emitting a single shock wave: (a)~High-speed shadowgraph visualization. The interframe time is 100~ns and the black bar shows the 1~mm scale. See supplementary movie \textit{Video\_Fig4.avi}. (b)~Pressure recorded by the hydrophone. The inset shows the whole bubble oscillation, where the orange and blue circles refer to generation and collapse shock wave peaks pressures, respectively. The dashed line shows $p(t)-p_{0}$ where $p(t)$ is the Rayleigh pressure model computed from Eq.~(\ref{eq:pre}) up to the shock peak, and the dotted line extends the curve to the time at which the bubble is estimated to reach a radius of $R=100$~$\mu$m.}
\label{fig:spherical}
\end{figure}

A spherical bubble collapse emits a single shock front that is spherically symmetrical, as visualized in Figure~\ref{fig:spherical_visu}.
This shock is well studied and arises from the compression of the incondensable gases inside the bubble overcoming the high pressures in the liquid around the bubble in the final collapse stage, which makes the liquid rapidly invert its motion as the bubble rebounds~\cite{Hickling1964}.
The gases inside the bubble are compressed so violently that they heat up to temperatures reaching levels of visible light emission, a phenomenon known as luminescence, which is visible in frame 5 of Figure~\ref{fig:spherical_visu} and implies that the bubble reaches its minimum size during the 50~ns exposure time of this image.
The rebound bubble then forms a compression wave that propagates outwards and quickly steepens to form a shock front, as seen in frames 6-8.
The corresponding hydrophone measurement of the shock wave is shown in Figure~\ref{fig:spherical_hydro}. 
Assuming $1/r$ spreading of the spherical wave and the luminescence spot in Figure~\ref{fig:spherical_visu} as the minimum bubble size ($R_{\rm min}\approx100$~$\mu$m), the lower bound for the peak pressure at the bubble wall at minimum bubble radius is estimated as $2$~GPa which is in agreement to previously estimated values~\cite{Lauterborn2013}. 
The actual value is likely much higher, because we overestimate the minimum bubble radius that our apparatus is not able to capture due to the luminescence and the dark region around the bubble hiding this information. 
When using the Keller-Miksis model~\cite{Keller1980}, where we adjust the initial gas pressure by numerically fitting the model to the observed radial evolution of the bubble (first and second oscillation), we would expect a minimum bubble radius of $R_{\rm min}\approx15~\mu$m, and thereby a peak pressure of  $12$~GPa.

In agreement with previous research, we find that the most energetic shock waves are emitted by highly spherical collapses, reaching up to about 90\% of the initial bubble energy.
The bubbles here are found to emit a single shock front at anisotropies up to $\zeta \approx 10^{-3}$ (equivalent to $\gamma \approx 14$), which is also the approximate limit for the appearance of a micro-jet piercing the bubble in our setup~\cite{Supponen2016}.

In the last stages of the collapse, the pressure in the liquid near the bubble wall increases to values so high that it deflects light, producing the shaded ring around the bubble in Figure~\ref{fig:spherical_visu} (frames 2-4).
This pressure has previously been predicted to reach thousands of bars~\cite{Rayleigh1917,Hickling1964} and experimentally detected using Mach-Zehnder interferometry~\cite{Ward1991} or elevated ambient pressures~\cite{Sukovich2017}.
However, it is interesting that our setup is able to visualize it using simple shadowgraphy at atmospheric pressure.
This is due to the bubble's high initial sphericity allowing it to reach very small radii upon its exceptionally spherical collapse.

The incompressible model for the pressure distribution around the bubble, developed by Rayleigh a century ago, is given as follows~\cite{Rayleigh1917}:
\begin{equation}
\frac{p}{p_{0}} = 1+\frac{R}{3r}\left(\frac{R_{0}^{3}}{R^{3}}-4\right)-\frac{R^{4}}{3r^{4}}\left(\frac{R_{0}^{3}}{R^{3}}-1\right)
\label{eq:pre}
\end{equation}
where $r$ is the radial distance from the bubble center.
Considering the lower bound for the compression ratio of the bubble in Figure~\ref{fig:spherical_visu} ($R_{0}/R_{\rm min} > 40$), we expect the maximum peak pressure to be in the order of GPa in the incompressible framework. 
The pressure buildup is visible in the hydrophone signal in Figure~\ref{fig:spherical_hydro} as a relatively slow rise preceding the peak pressure of the shock.
We may compute the pressure evolution in time from Eq.~(\ref{eq:pre}) at the radial distance where the hydrophone is located ($r=44.5$~mm), assuming the time evolution of the bubble radius to follow the analytical approximation $R(t)\approx R_{0}\left(1-t^{2}\right)^{2/5}$~\cite{Obreschkow2012} (where $t$ is the time normalized to collapse time $T_{c}$), down to $R_{\rm min}\approx 100$~$\mu$m.
The computed pressures from Eq.~(\ref{eq:pre}) can be roughly compared with the hydrophone signal if the delay in the far-field caused by the finite sound speed is accounted for.
Furthermore, the shock pressure peak is assumed to represent a time approximately 100~ns preceding the final collapse instant, for the shock wave is expected to propagate the first $\sim 300$~$\mu$m with supersonic speeds~\cite{Vogel1996}.
The average shock speed during the exposure of the first frame after the collapse is estimated approximately as 3000~$ms^{-1}$ from Figure~\ref{fig:spherical_visu}, and therefore the shock wave is indeed estimated to reach the hydrophone $\Delta t \approx102$~ns earlier than the pressure buildup, of which the information is assumed to propagate at the sound speed.
As seen in Figure~\ref{fig:spherical_hydro}, the computed (dashed line) and measured (solid line) pressure evolutions almost up to the signal peak are surprisingly similar.
The good agreement is remarkable considering our unconventional pressure calibration.
The model is not able to reproduce the shock wave because it is incompressible (dotted line), and when the bubble reaches a radius of $R=100$~$\mu$m, the predicted pressure at the hydrophone location is $p-p_{0}=3.8$~MPa, which is close to the measured peak pressure very likely by coincidence.
The pressure rise, in addition to the tensile part of the shock wave tail, is the clearest difference between the measured waveform from a spherical collapse and that of the bubble generation (Figure~\ref{fig:gen}).

\subsection{Non-spherical collapse: Bubbles near a free surface}

\begin{figure}[b]
\begin{center}
\includegraphics[width=\textwidth]{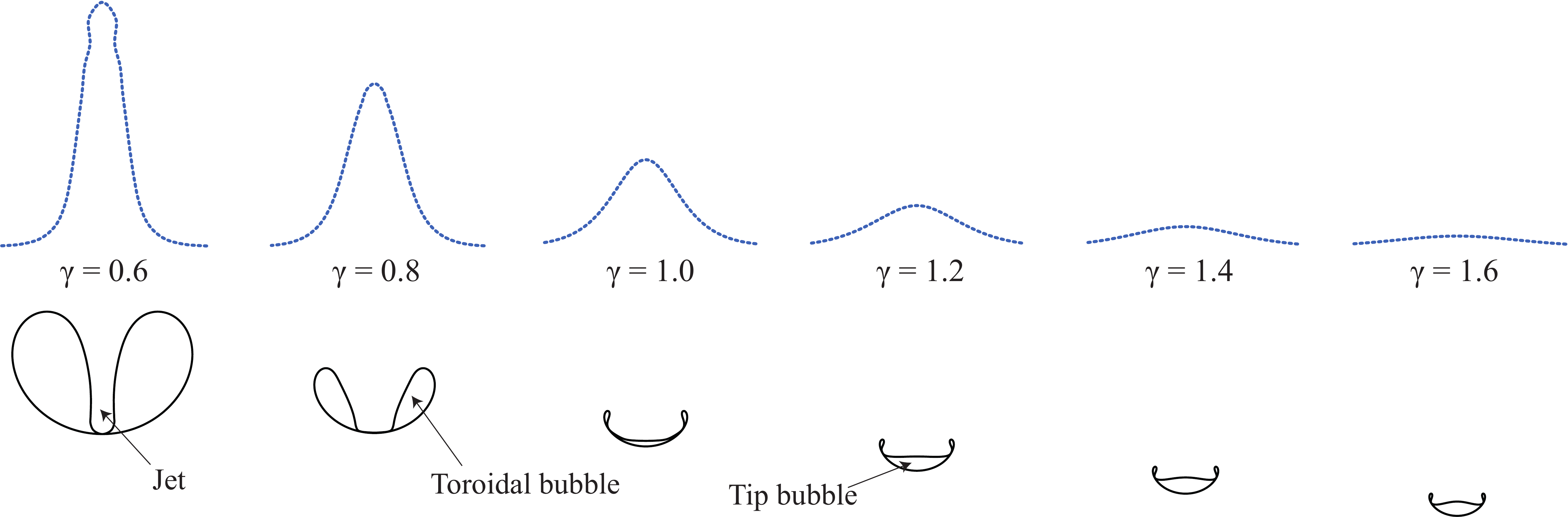}
\caption{Illustration of the bubble shapes at jet impact for different stand-off distances $\gamma$ from the free surface. The corresponding values for $\zeta$ from left to right are $\zeta=0.54$, 0.30, 0.20, 0.14, 0.10 and 0.076. The shapes of the free surface are shown as a dotted line. The shapes have been obtained numerically using potential flow theory.}
\label{fig:freeshapes}
\end{center}
\end{figure}

The dynamics of bubbles near free surfaces has been extensively studied in the past experimentally, theoretically and numerically~\cite{Chahine1977,Blake1981,Blake1987,Wang1996,Robinson2001,Pearson2004,Zhang2016,Koukouvinis2016}, yet no study to-date has focused specifically on their shock wave emission.
The advantage of studying bubbles near a free surface is the contact avoidance between the bubble and the surface, allowing thus free collapse and rebound dynamics, as the bubble migration and the micro-jet are directed away from the surface (in contrary to a rigid surface).
While bubbles near a free surface form micro-jets that have similar characteristics to bubbles deformed by a rigid surface~\cite{Supponen2016}, their shapes at the final collapse stages have significant differences, which may give us some further insight into the distinct shock wave emission mechanisms.
In particular, for $\gamma=1$--3, the micro-jet formed during the collapse is broad and impacts the opposite bubble wall on a ring rather than a single point, some examples being illustrated in Figure~\ref{fig:freeshapes}.
At lower values of $\gamma$, the micro-jet becomes narrow and the spike formed on the free surface increases in height.
The shapes in Figure~\ref{fig:freeshapes} were obtained numerically using potential flow theory (boundary integral method~\cite{Supponen2016,Taib1983,Blake1987,Robinson2001}\footnote{The code for the numerical simulations is available online at \url{https://obreschkow.shinyapps.io/bubbles}~\cite{Supponen2016}.}) and have previously been validated by their good agreement with experiments~\cite{Supponen2016}.

We now present observations of shock waves from bubbles collapsing near a free surface at different levels of $\zeta$.
Non-spherically collapsing bubbles that produce micro-jets generate multiple shock waves, which are clearly observed on the shadowgraph images at $\zeta > 10^{-3}$. 
However, they only become clearly distinct events on the hydrophone signal beyond $\zeta \sim 8\times10^{-3}$ ($\gamma \sim 5$).

\begin{figure}
\begin{center}
\begin{subfigure}{0.4\textwidth}
	\begin{overpic}[width=\textwidth]{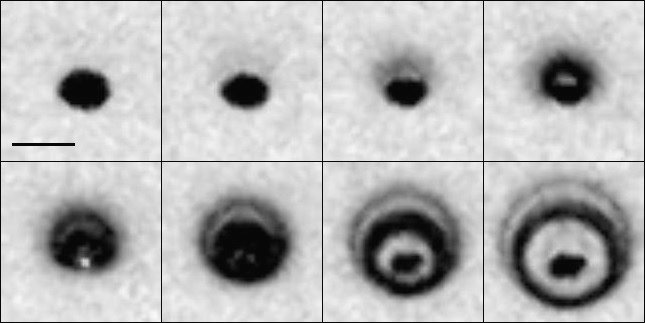}
	\end{overpic}
	\caption{ }
	\label{fig:weak_visu2}
\end{subfigure}
\begin{subfigure}{0.58\textwidth}
	\includegraphics[width=\textwidth]{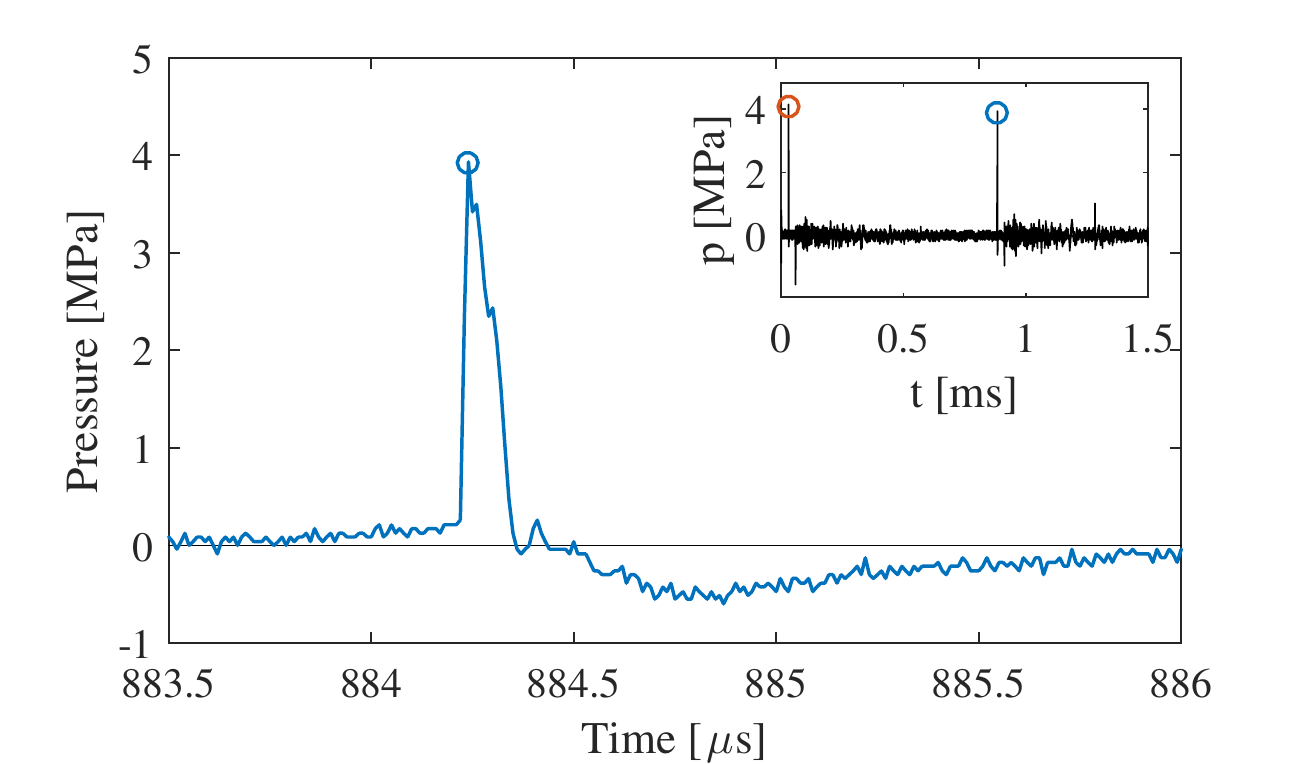}
	\caption{ }
	\label{fig:weak_hydro2}
\end{subfigure}
\end{center}
\caption{Cavity of $R_{0} = 4.1$~mm at $\zeta = 3.8\times10^{-3}$ ($\gamma = 7.2$): (a)~High-speed shadowgraph visualization. The interframe time is 100~ns and the black bar shows the 1~mm scale. See supplementary movie \textit{Video\_Fig6.avi} (b)~Pressure recorded by the hydrophone. The inset shows the whole bubble oscillation, where the orange and blue circles refer to generation and collapse shock wave peak pressures, respectively.}
\label{fig:intermediate_lum}
\end{figure}

Figure~\ref{fig:intermediate_lum} shows selected shadowgraph images and the corresponding hydrophone pressures for a bubble collapsing at $\zeta = 3.8\times10^{-3}$.
The first sign of asymmetry in the bubble collapse, together with the bubble's displacement, is the shaded region appearing near the upper bubble wall where the downward micro-jet is forming (starting from frame 2 in Figure~\ref{fig:weak_visu2}).
It is similar to the gradual pressure buildup observed for the spherical collapse in Figure~\ref{fig:spherical_visu}, but not spherically symmetric.
It is also in agreement with reported numerical simulations of jetting bubbles, finding higher pressures at the root of the jet relative to the rest of the pressure field~\cite{Chahine2015,Koukouvinis2016,Koukouvinis2016b,Li2016}.
The shaded region eventually surrounds most of the bubble in frame 5, and two clear shock fronts are visible in frame 6 following the collapse.
We observe luminescence at the tip of the bubble in frame 5, which also appears to be the center of the most pronounced shock wave visible in the subsequent images.
Although it is much weaker compared to the light emitted in the spherical case, the observed flash suggests a high gas compression between the jet and the opposite bubble wall.
Interestingly, the first shock front in Figure~\ref{fig:weak_visu2} is produced on the side of the bubble where the initial pressure rise in the liquid occurred. 
The hydrophone is unable to distinguish the first shock wave from the rest due to its location and temporal resolution, but it records the gradual pressure rise occurring on the sides of the bubble preceding the main shock wave (Figure~\ref{fig:weak_hydro2}).

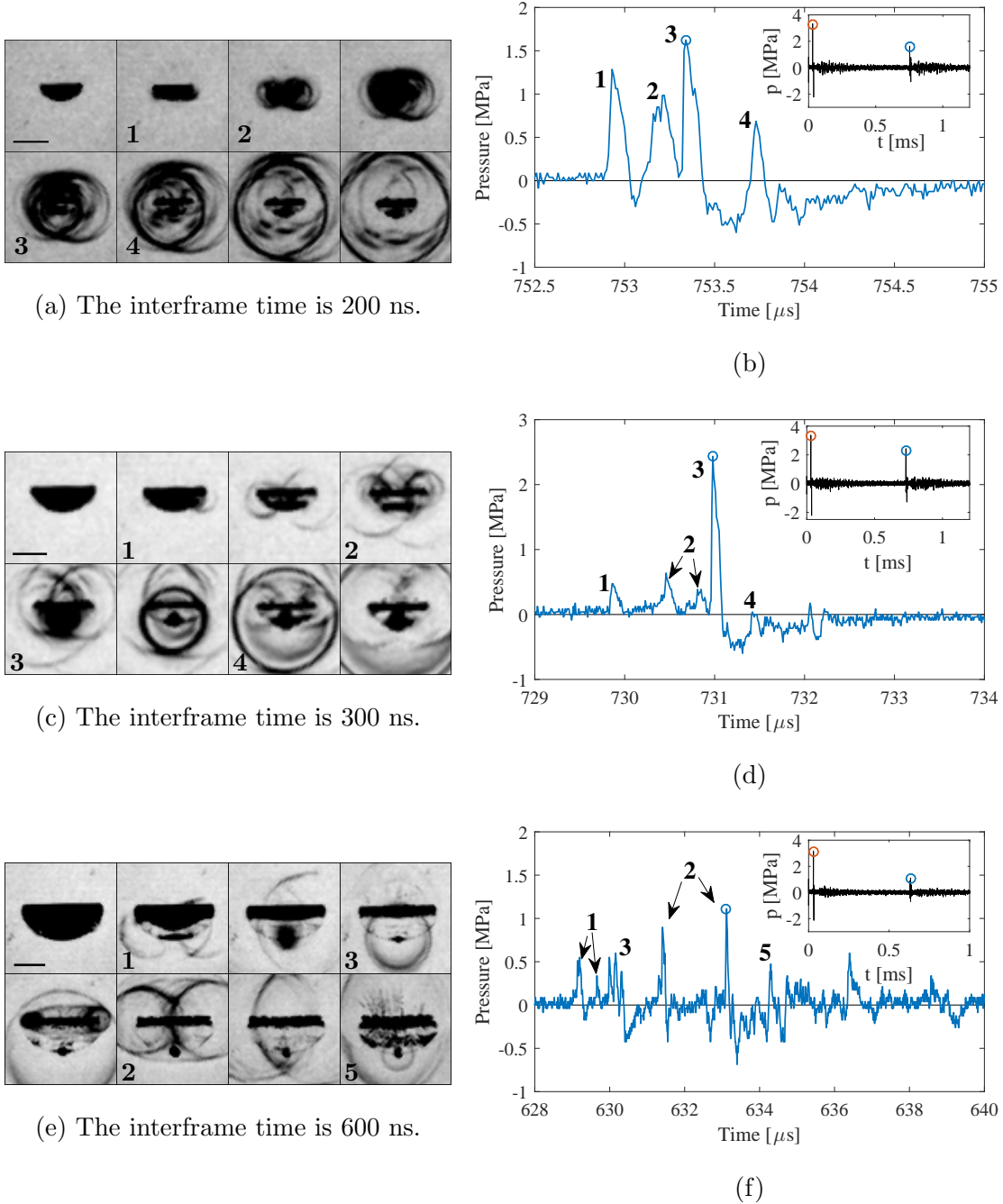
\begin{figure}[b]
\begin{center}
\begin{subfigure}{0.4\textwidth}
	\begin{overpic}[width=\textwidth]{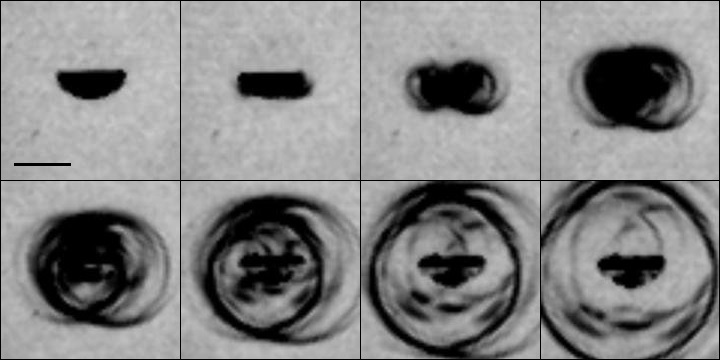}
	\put (27,27) {\textbf{1}}
	\put (52,27) {\textbf{2}}
	\put (2,2) {\textbf{3}}
	\put (27,2) {\textbf{4}}
	\end{overpic}
	\caption{The interframe time is 200~ns.}
	\label{fig:jet_visu}
\end{subfigure}
\begin{subfigure}{0.52\textwidth}
	\includegraphics[width=\textwidth]{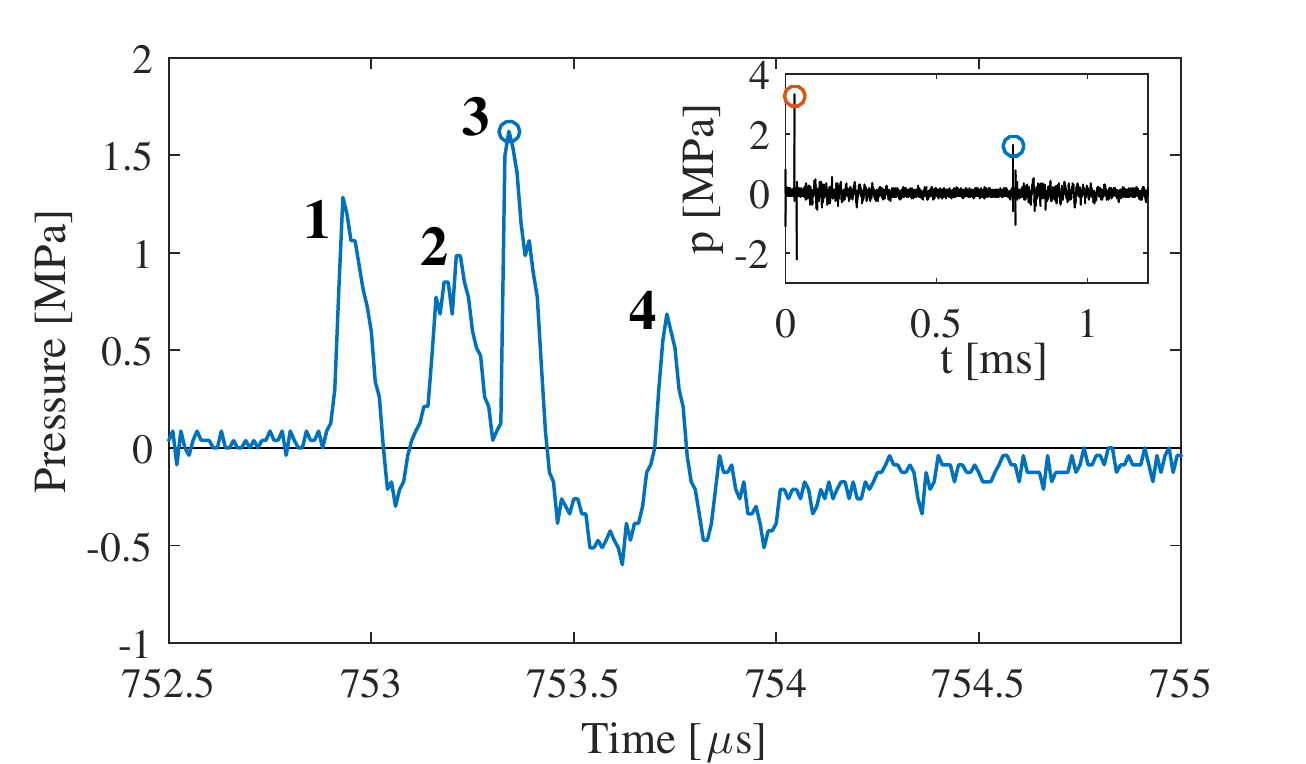}
	\caption{ }
	\label{fig:jet_hydro}
\end{subfigure}
\begin{subfigure}{0.4\textwidth}
	\begin{overpic}[width=\textwidth]{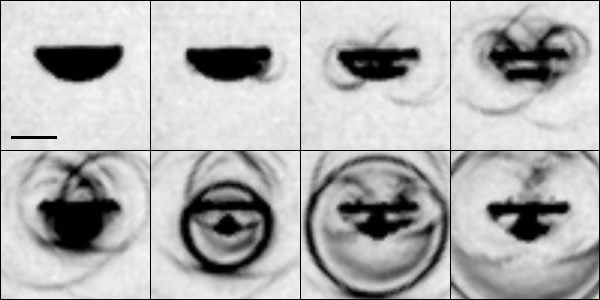}
	\put (26,26) {\textbf{1}}
	\put (76,26) {\textbf{2}}
	\put (1,1) {\textbf{3}}
	\put (51,1) {\textbf{4}}
	\end{overpic}
	\caption{The interframe time is 300~ns.}
	\label{fig:tip_visu}
\end{subfigure}
\begin{subfigure}{0.52\textwidth}
	\includegraphics[width=\textwidth]{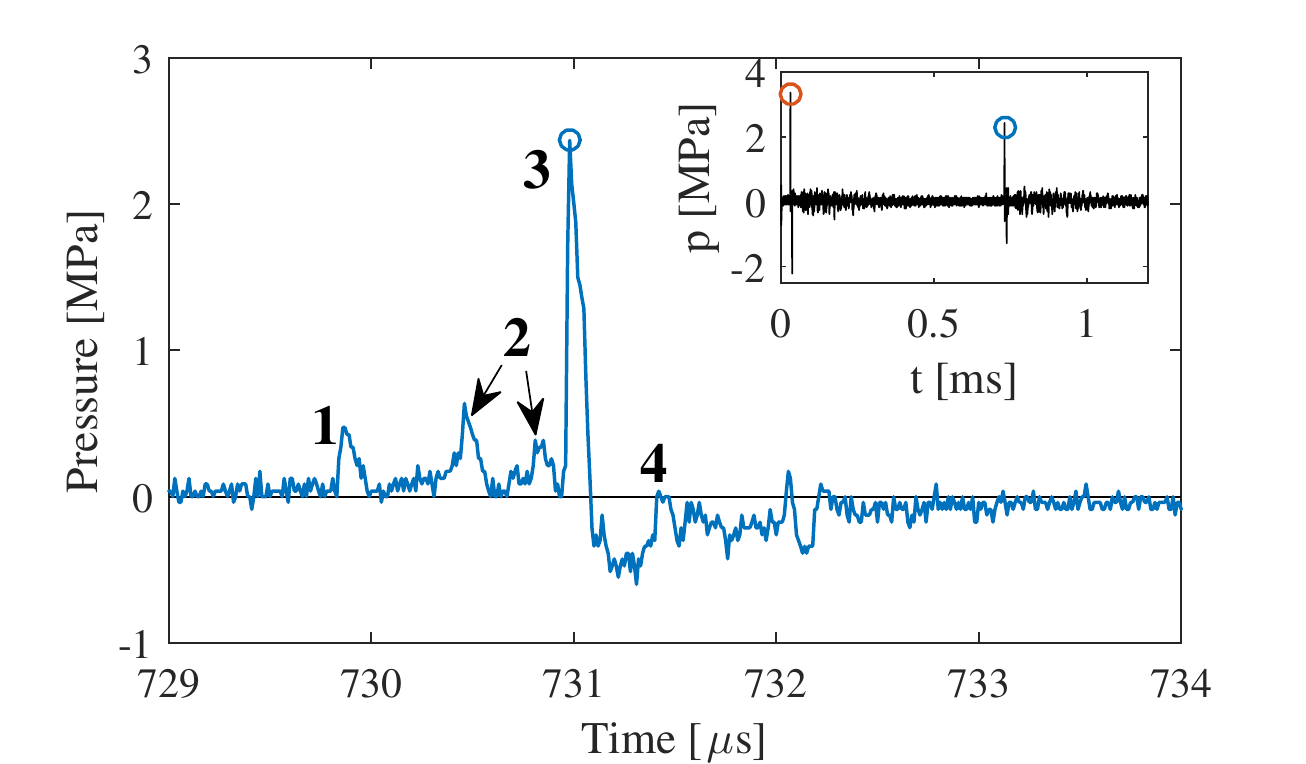}
	\caption{ }
	\label{fig:tip_hydro}
\end{subfigure}
\begin{subfigure}{0.4\textwidth}
	\begin{overpic}[width=\textwidth]{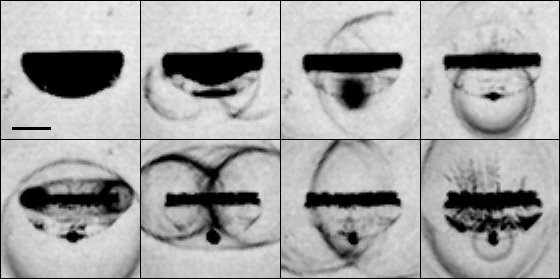}
	\put (26,26) {\textbf{1}}
	\put (76,26) {\textbf{3}}
	\put (26,1) {\textbf{2}}
	\put (76,1) {\textbf{5}}
	\end{overpic}
	\caption{The interframe time is 600~ns.}
	\label{fig:tip_visu2}
\end{subfigure}
\begin{subfigure}{0.52\textwidth}
	\includegraphics[width=\textwidth]{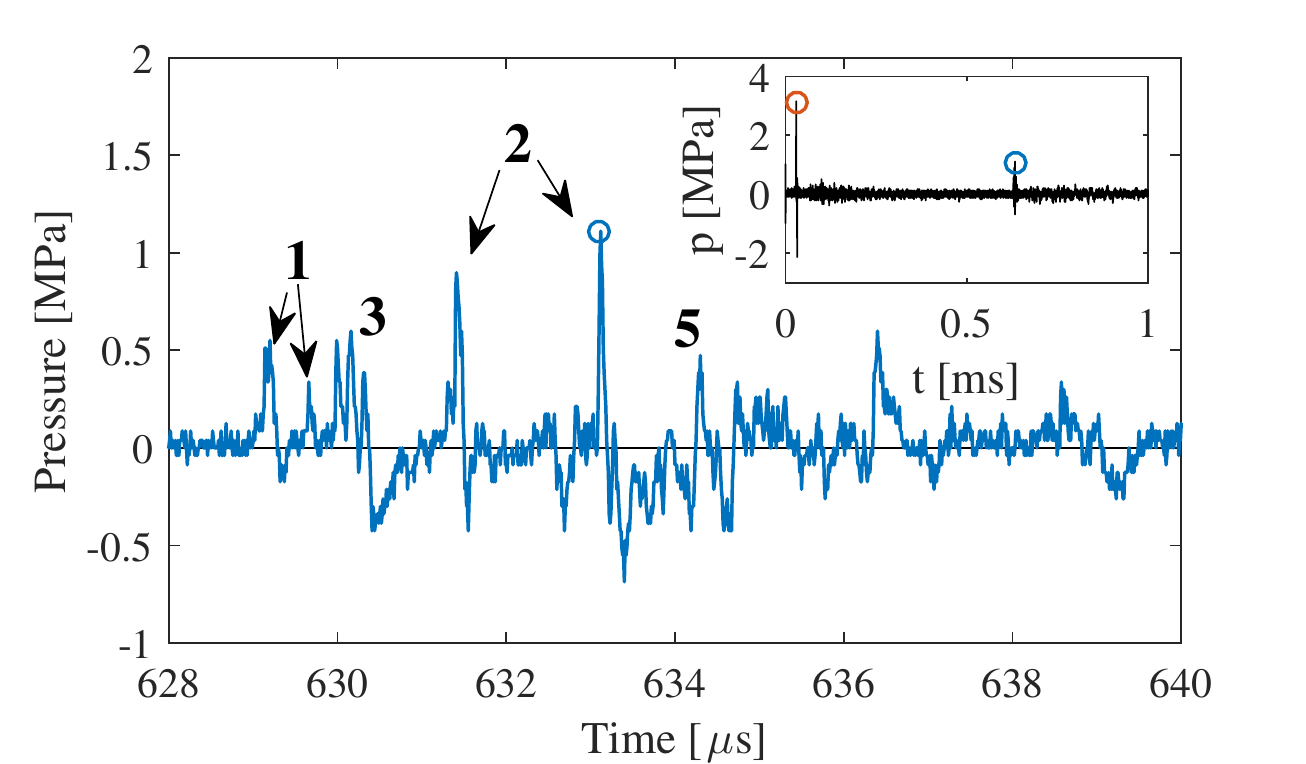}
	\caption{ }
	\label{fig:tip2_hydro}
\end{subfigure}
\end{center}
\caption{Caption on next page}
\label{fig:intermediate1}
\end{figure}
\begin{figure}
\begin{center}
    \ContinuedFloat
    \captionsetup{list=off}
\begin{subfigure}{0.4\textwidth}
	\begin{overpic}[width=\textwidth]{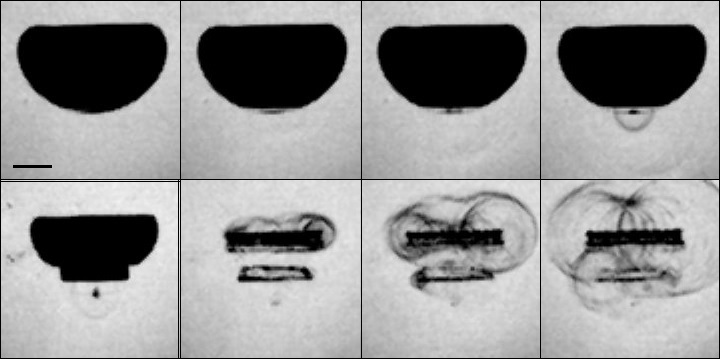}
	\put (26,26) {\textbf{1}}
	\put (76,26) {\textbf{3}}
	\put (1,1) {\textbf{5}}
	\put (51,1) {\textbf{2}}
	\put (0,20) {+15~$\mu$s}
	\put (25,20) {+13~$\mu$s}
	\end{overpic}
	\caption{The interframe time is 400~ns unless otherwise indicated.}
	\label{fig:torus_visu}
\end{subfigure}
\begin{subfigure}{0.52\textwidth}
	\includegraphics[width=\textwidth]{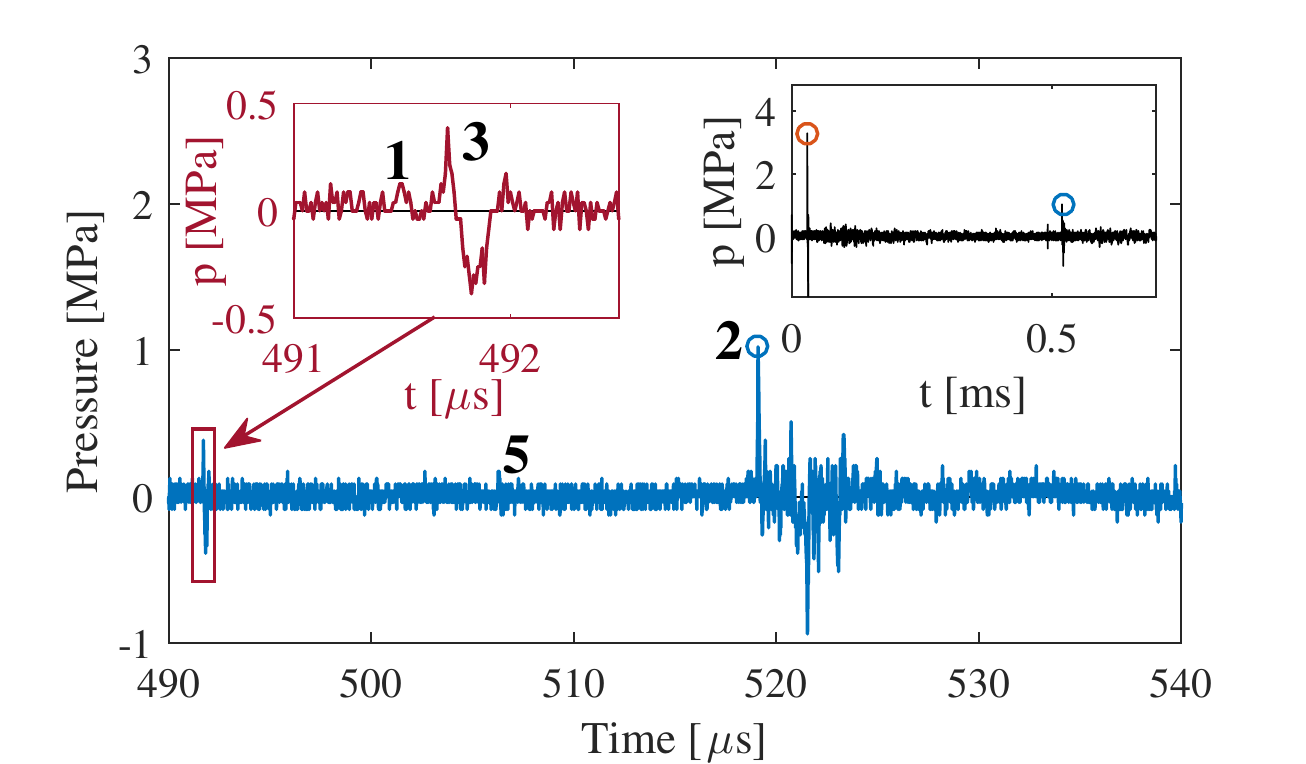}
	\caption{ }
	\label{fig:torus_hydro}
\end{subfigure}
\end{center}
\caption{(Continued) Selected images (left) and hydrophone signal (right) for cavities of (a),(b) $R_{0} = 3.6$~mm at  $\zeta = 2.9\times10^{-2}$ ($\gamma = 2.6$), (c),(d) $R_{0} = 3.6$~mm at $4.6\times10^{-2}$ ($\gamma = 2.1$), (e),(f) $R_{0} = 3.2$~mm at $\zeta = 0.19$ ($\gamma = 1$), and (g),(h) $R_{0} = 3.0$~mm at $\zeta = 0.33$ ($\gamma = 0.77$). The shock waves are denoted as: 1) jet impact, 2) torus collapse, 3) tip bubble collapse, 4) second torus collapse and 5) second tip bubble collapse shock waves. The black bars show the 1~mm scale. The insets show the whole bubble oscillation, where the orange and blue circles refer to generation and collapse shock wave peak pressures, respectively. See supplementary movies \textit{Video\_Fig7a.avi}, \textit{Video\_Fig7c.avi}, \textit{Video\_Fig7e.avi} and \textit{Video\_Fig7g.avi} (\textit{Video\_Fig7g.avi} combines films made of two separate bubbles due to the long duration of the events and the limited number of frames captured by the camera. The events are highly repetitive.)}
\label{fig:intermediate1}
\end{figure}

Figures~\ref{fig:jet_visu}--\ref{fig:torus_hydro} show images and the corresponding measured shock pressures for more deformed bubbles, collapsing at different distances from the free surface at $\zeta = 2.9\times10^{-2}$, $4.6\times10^{-2}$, 0.19 and 0.33.
The recorded peak pressures are significantly lower compared to the more spherical cases, and many distinct shock wave events are observed.
The first pressure peak in all cases corresponds to the water hammer induced by the jet impact.
Such a shock has been observed in the past for non-spherically collapsing bubbles both experimentally~\cite{Ohl1999,Lindau2003,Supponen2015} and numerically~\cite{Johnsen2009,Koukouvinis2016}.
It produces a torus-like shock wave due its contact on the opposite bubble wall not being a single point but a circular line (see Figure~\ref{fig:freeshapes}), clearly visible on the images as two shock source `points' on the sides of the bubble.
If the jet is broad enough, the hydrophone may detect two individual pressure peaks, such as in Figure~\ref{fig:tip2_hydro}, owing to such torus-like shock having two fronts on the hydrophone axis that reach the sensor.
Subsequently, the jet separates a part of the vapor at the `tip' from the rest of the bubble, with this separation being particularly clear in Figures~\ref{fig:tip_visu} and \ref{fig:tip_visu2} as a horizontal line that cuts the bubble and implies that the vapor in that zone has disappeared.
It is difficult to tell with certainty that the first shock wave results from a jet impact in Figure~\ref{fig:jet_visu} due to the short time intervals between the distinct events.
However, observing several bubbles between $\zeta=2.9\times10^{-2}$ and $4.6\times10^{-2}$ (of which the results are summarized later in Section~\ref{s:dis}), a systematic variation of the shock timings and strengths with $\zeta$ was noted.
The identification of each peak in Figure~\ref{fig:jet_hydro} was therefore done accordingly.
The peak pressure associated with the jet impact decreases with an increasing $\zeta$, and is barely detected at $\zeta =0.33$.
At $\zeta = 2.9\times10^{-2}$ and $4.6\times10^{-2}$ (Figures \ref{fig:jet_visu}--\ref{fig:tip_hydro}), the jet impact is followed by the collapse of the toroidal bubble.
The associated shocks are torus-like and meet in the jet axis in the middle of the bubble, which is known to sometimes produce a `counter-jet', a vertical column-like cluster of micro-cavities~\cite{Lindau2003,Supponen2016}.
The torus collapse shock may also yield two individual peaks in the pressure signal, such as in Figures~\ref{fig:tip_hydro} and \ref{fig:tip2_hydro}.
The peak pressure of the torus collapse shock first decreases with increasing $\zeta$ (Figures~\ref{fig:jet_hydro} and \ref{fig:tip_hydro}), and then increases again slightly (Figures~\ref{fig:tip2_hydro} and \ref{fig:torus_hydro}).
The next pressure peak in Figures~\ref{fig:jet_hydro}~and~\ref{fig:tip_hydro} corresponds to the tip bubble collapse.
It appears to be the dominant shock in the collapse scenario at these $\zeta$.
The tip bubble collapse shock triggers a second collapse of the rebounding toroidal bubble, which emits a further shock wave manifested as the fourth pressure peak in the signal.
The second torus collapse pressure peak is considerable at $\zeta = 2.9\times10^{-2}$ but barely detected by the hydrophone at $\zeta = 4.6\times10^{-2}$.
As seen in Figures~\ref{fig:tip_visu2}~and~\ref{fig:torus_visu}, at a higher $\zeta$ the tip bubble collapse and the torus collapse change order.
In Figure~\ref{fig:torus_visu} the tip bubble is very small and its collapse follows the jet impact so closely that it is difficult to distinguish the shocks they emit.
At $\zeta = 0.19$ it is the torus collapse that triggers a second collapse of the tip bubble, while at $\zeta = 0.33$ the tip bubble is able to collapse naturally a second time long before the torus collapse.
In Figure~\ref{fig:torus_visu}, the compression of the toroidal bubble is highly non-uniform, yielding multiple peaks that generate a noisy hydrophone signal (Figure~\ref{fig:torus_hydro}).

\begin{figure}
\begin{center}
\begin{subfigure}{0.33\textwidth}
	\begin{overpic}[width=\textwidth]{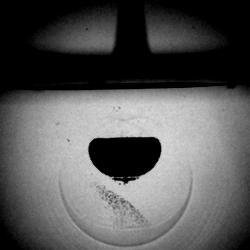}
	\put (2,93) {\textcolor{white}{$t=0.46$~ms}}
	\put (29,17) {\textbf{1}}
	\put (35,17) {$\boldsymbol{\rightarrow}$}
	\put (19,5) {\textbf{2}}
	\put (25,5) {$\boldsymbol{\rightarrow}$}
	\put (63,19) {\textbf{3}}
	\put (63,11) {$\boldsymbol{\downarrow}$}
	\end{overpic}
	\caption{ }
	\label{fig:Sec_cav1}
\end{subfigure}
\begin{subfigure}{0.33\textwidth}
	\begin{overpic}[width=\textwidth]{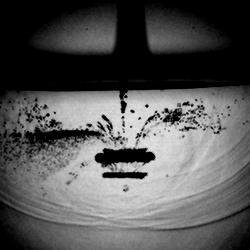}
	\put (2,93) {\textcolor{white}{$t=0.49$~ms}}
	\put (33,56) {\textbf{3}}
	\put (39,56) {$\boldsymbol{\rightarrow}$}
	\put (69,33) {\textbf{2}}
	\put (61,39) {$\boldsymbol{\nwarrow}$}
	\put (22,19) {\textbf{1}}
	\put (22,28) {$\boldsymbol{\uparrow}$}
	\put (45,1) {\textbf{4}}
	\put (35,3) {$\boldsymbol{\nwarrow}$}
	\put (51,3) {$\boldsymbol{\nearrow}$}
	\end{overpic}
	\caption{ }
	\label{fig:Sec_cav2}
\end{subfigure}
\end{center}
\caption{Visualization of secondary cavitation resulting from the passage of rarefaction waves for the same bubble as in Figure~\ref{fig:torus_visu} at two different instants: (a)~Secondary cavitation (1) below the bubble, generated by the tip bubble collapse shock wave (2) turned into a rarefaction wave (3) after reflecting at the bubble's interface; and (b)~Secondary cavitation visible in the pre-heated cone-shaped zone in the laser path (1), as streamers along the micro-jet flow (2) and as a vertical column (3), generated by the rarefaction waves (4) caused by the reflection of torus collapse shock waves at the free surface. These are selected images from the supplementary movie \textit{Video\_Fig7g.avi}.}
\label{fig:secondary}
\end{figure}

The shock wave strengths are also visible as the darkness levels of the corresponding image pixels owing to their ability to deflect light, which is seen, for example, in Figure~\ref{fig:tip_visu} where the tip bubble shock wave is clearly the most pronounced of all the events.
The time intervals between each event substantially increase with $\zeta$.
When the bubble collapses very close to the free surface, the hydrophone also detects the reflected rarefaction waves following closely the original shocks and contributing to the noise in the signal of Figure~\ref{fig:torus_hydro}.
These waves are visible in all movies of Figure~\ref{fig:intermediate1} and, due to their negative pressure resulting from the reflection at the free surface, they generate secondary cavitation in the bubble's neighborhood, as shown in Figure~\ref{fig:secondary}.
The secondary cavities are visible as clusters of micro-bubbles most prominently in the path of the focused laser, where the liquid is pre-heated and thereby the nucleation of cavities is facilitated, and between the bubble and the free surface (Figure~\ref{fig:Sec_cav2}).
Interestingly, some of these clusters, arranged in streamers towards the central axis of the toroidal bubble, delineate the flow induced by the formation of the micro-jet.
The vertical column of micro-bubbles between the toroidal bubble and the free surface in Figure~\ref{fig:Sec_cav2} appears to result from the confluence of the rarefaction waves that are the reflections of the shocks initially emitted by the torus collapse.
For the same bubble, secondary cavitation resulting from the shock emitted at the first tip bubble collapse is also observed below the bubble, right after the jet impact, as seen in Figure~\ref{fig:Sec_cav1}.
Here the negative pressure results from the reflection at the bubble interface, and the rarefaction wave follows closely the original shock wave, which explains the significant tensile tail of the tip bubble collapse peak captured by the hydrophone in Figure~\ref{fig:torus_hydro}.

\subsection{Energy distribution and event timings}
\label{s:dis}

The observations of the distinct shock wave events and their corresponding pressures show important variations with different bubble asymmetries.
The energy of the observed shock waves can be estimated from the hydrophone pressure signal via Eq.~(\ref{eq:ES}), where the integration range is selected by identifying the pressures associated to each individual event from the high-speed visualizations.
It should be noted that this method assumes spherically symmetric propagation of the shock wave.
Some shocks, especially the jet impact shock, might have some directionality, biasing their energy measurement. 
Indeed, it has been shown numerically that jet impact-induced shocks are dependent on the orientation with respect to the jet close to the bubble~\cite{Johnsen2009,Hsiao2014}.
However, the symmetric shock shadings seen in the high-speed visualizations far from the bubble center (not shown in figures) suggest that this directionality must be sub-dominant. 
The shock pressure dependence on orientation likely reduces as the wave propagates and decreases in amplitude.
We nonetheless caution that directionality is a potential source of systematic error, which might be reduced in future experiments by using multiple hydrophones in different directions.

\begin{figure}
\begin{center}
\includegraphics[width=0.8\textwidth]{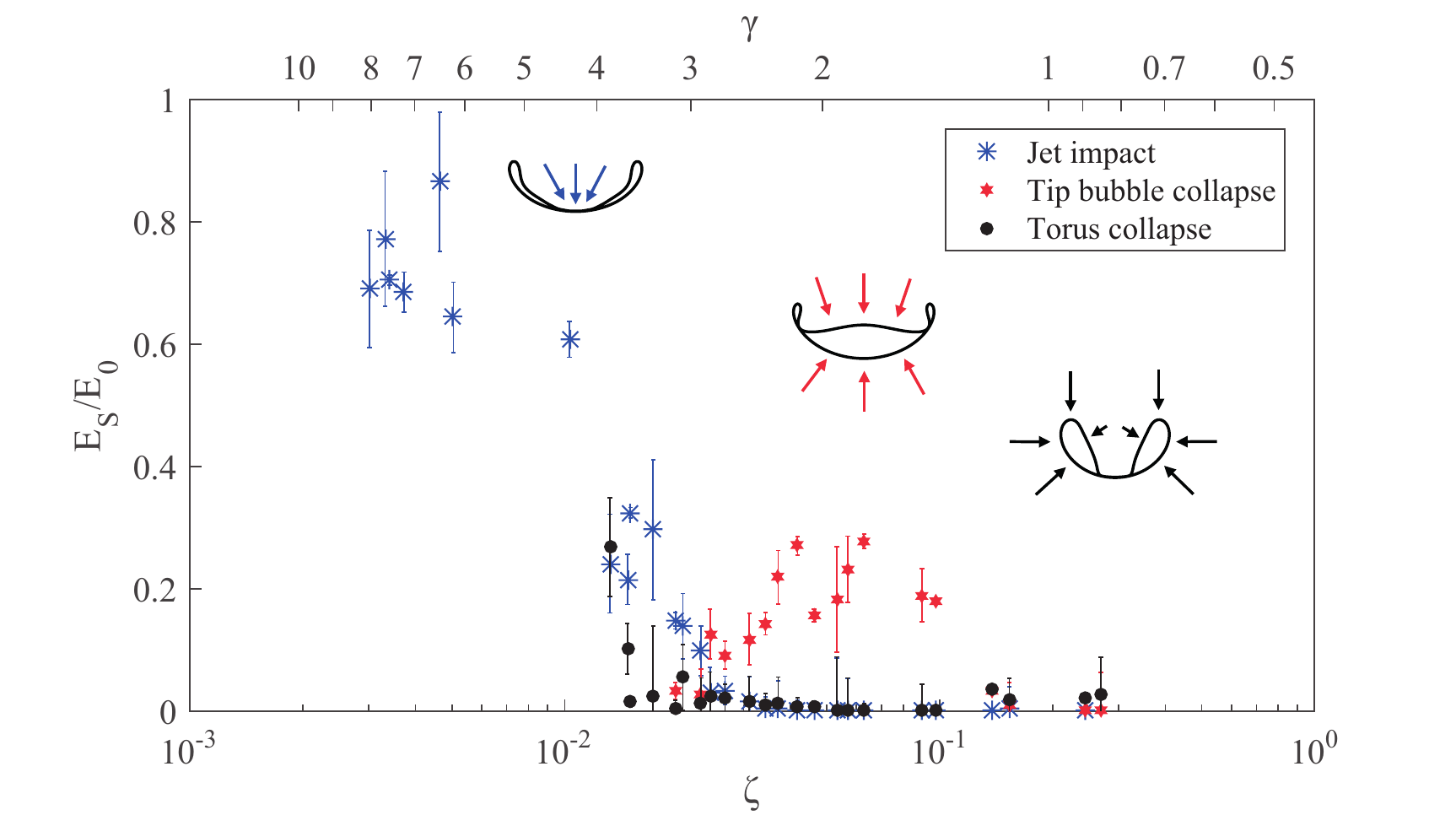}
\end{center}
\caption{Normalized shock wave energy for each shock emission mechanism from bubbles deformed by a near free surface, as a function of $\zeta$ (and corresponding $\gamma$, top axis). Numerically calculated bubble shapes at jet impact are shown for $\zeta= 10^{-2}$, $6\times10^{-2}$ and 0.3.}
\label{fig:z_E_all}
\end{figure}

The fraction of the bubble's initial energy $E_{0}$ distributed to the distinct shock waves for bubbles collapsing near a free surface is shown in Figure~\ref{fig:z_E_all} as a function of the anisotropy parameter $\zeta$ (and the equivalent $\gamma$).
We only measured bubbles up to $\zeta\sim0.3$ ($\gamma \sim 0.8$), beyond which the free surface resulted in severe perturbations in the hydrophone signal due to the reflected rarefaction waves.
The driving pressure was kept $\Delta p > 75$~kPa in order to avoid simultaneous deformations by the free surface and gravity, which could lead to more complex shapes at the bubble collapse (e.g.~bubble splitting or annular jets~\cite{Cui2016}).
The energy of each of the three main shock waves, i.e.\ jet impact, tip bubble collapse and torus collapse, vary as functions of $\zeta$.
Interestingly, each of them dominates a certain range of $\zeta$, as seen in Figure~\ref{fig:z_E_all}.
For bubbles that produce jets, the jet impact shock appears to dominate up to $\zeta \sim 2\times10^{-2}$.
The tip bubble shock wave has a clear domination in the range $2\times10^{-2} < \zeta < 0.15$.
Beyond $\zeta \sim 0.15$, the torus collapse shock wave is the most energetic, yet weak in relative terms with less than $10\%$ of the initial bubble energy.
The torus collapse energy is particularly low in the range $2\times10^{-2}<\zeta < 0.1$, coinciding with the domination of the tip bubble.
The second torus collapse and the second tip bubble collapse emit shock waves with a negligible energy compared to the others, which is why they have been excluded from the figures.

The domination of the tip bubble in the range $2\times10^{-2} < \zeta < 0.15$ is explained through its large volume relative to the rest of the bubble at the moment of the jet impact, its spherical topology that allows an effective gas compression during its collapse, and/or the further compression provided by the pushing jet.
The large volume of the tip bubble and the small volume of the torus in this range result from the characteristic shape the jet assumes for bubbles collapsing near a free surface (see Figure~\ref{fig:freeshapes}).
Beyond $\zeta \sim 0.1$ however, the torus becomes relatively larger again at the moment of jet impact, as the bubble shape at $\zeta=0.3$ in Figure~\ref{fig:z_E_all} suggests, and the torus is able to compress the gases it contains more effectively.
This explains the slight rise of the torus collapse shock energy for $\zeta > 0.1$.

\begin{figure}
\begin{center}
\includegraphics[width=0.8\textwidth]{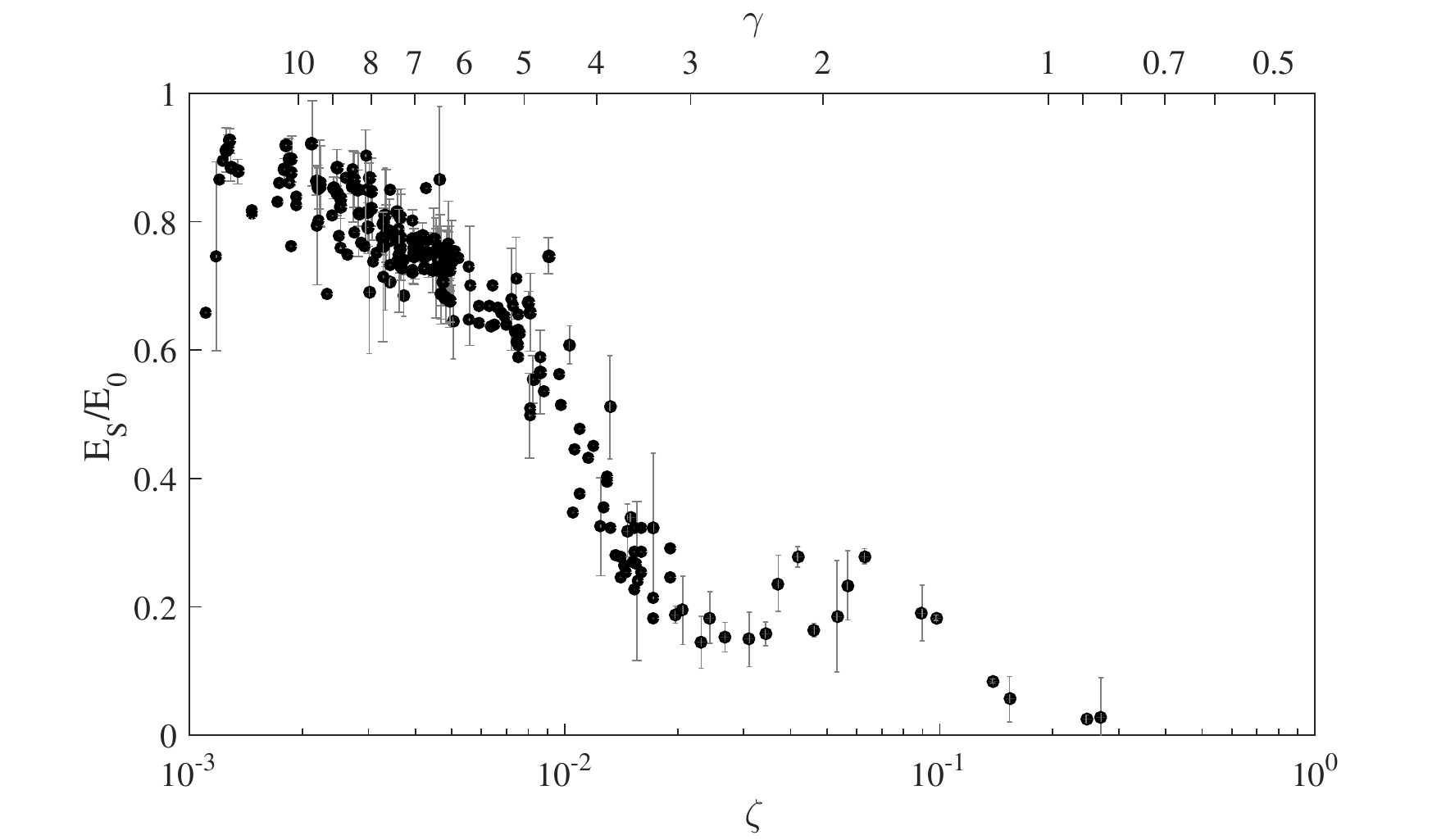}
\end{center}
\caption{Normalized total collapse shock wave energy $E_{S}/E_{0}$ for bubbles deformed by a near free surface, as a function of $\zeta$ (and $\gamma$, top axis).}
\label{fig:z_E_tot}
\end{figure}

When the energies of the different collapse shock waves are summed, an overall decrease of the total shock energy is observed, as seen in Figure~\ref{fig:z_E_tot}.
Here data for lower $\zeta$ have been added, including energies from pressure measurements for which it was not possible to distinguish the different shock wave events.
Interestingly, the total shock energy varies as a function of $\zeta$ independently of the bubble maximum radius and driving pressure within the ranges covered here ($R_{0}=1$--$4$~mm, $\Delta p = 0.75$--1~bar).
A major part of the collapse shock energy decrease occurs within the range $10^{-3}<\zeta<2\times10^{-2}$, where the jet impact hammer shock is expected to dominate.
As the bubble deforms, the liquid inflow towards the bubble center becomes anisotropic, and as a result, the level of compression of the bubble's enclosed gases reduces yielding weaker shock wave emission.
As less energy is radiated away by the shock waves for increasing $\zeta$, more energy is distributed to the motion of the liquid forming the micro-jet and to the rebound bubble, both of which are observed to grow with $\zeta$.

\begin{figure}
\begin{center}
\includegraphics[width=0.8\textwidth]{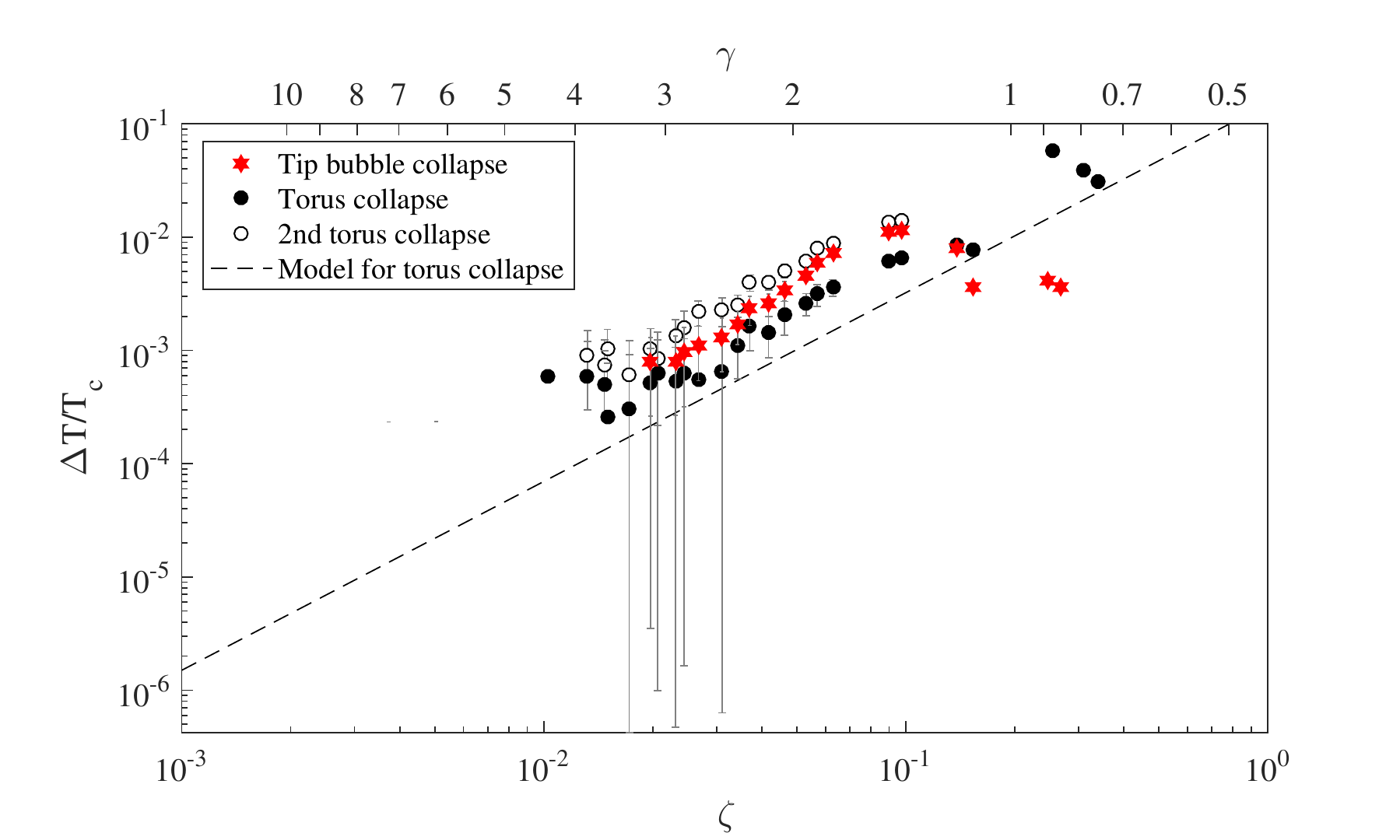}
\caption{Time differences between the jet impact and the tip bubble collapse, torus collapse and the second torus collapse as a function of $\zeta$ (and $\gamma$, top axis), normalized with bubble collapse time $T_{c}$. The time between jet impact and torus collapse is modeled as $\Delta T/T_{c} = 0.15\zeta^{5/3}$~\cite{Supponen2016}.}
\label{fig:z_deltaT}
\end{center}
\end{figure}

The timing of the distinct events in the shock wave scenario also appears to vary with the level of deformation of the bubble.
Figure~\ref{fig:z_deltaT} displays the time difference $\Delta T$ between the jet impact, that generally emits the first shock wave, and the other observed events, normalized to the bubble collapse time $T_{c}$.
The experiments are displayed together with our previously established model estimating the normalized time between the jet impact and torus collapse $\Delta T/T_{c} = 0.15\zeta^{5/3}$~\cite{Supponen2016}.
Only data for $\zeta>10^{-2}$ are displayed as the temporal resolution of our apparatus is not sufficient for identifying the exact shock timings of more spherical bubbles.
The jet impact occurs within the last 1\% of the bubble's collapse time up to $\zeta \approx 0.2$, followed very closely by the other events.
The torus collapse precedes the tip bubble collapse up to $\zeta \approx 0.14$, beyond which they change order.
The second torus collapse occurs right after the tip bubble collapse up to this limit, as the rebounding torus compresses under the effect of the shock wave produced by the latter, which is seen as an almost constant time difference between the two events in Figure~\ref{fig:z_deltaT}.
The normalized timings of each shock wave are independent of the maximum bubble radii and driving pressures covered here.

\section{Models for shock energy and pressure}
\label{s:models}

\begin{figure}
\begin{center}
\begin{overpic}[width=0.55\textwidth, trim=0cm 1.93cm 0cm 0cm, clip]{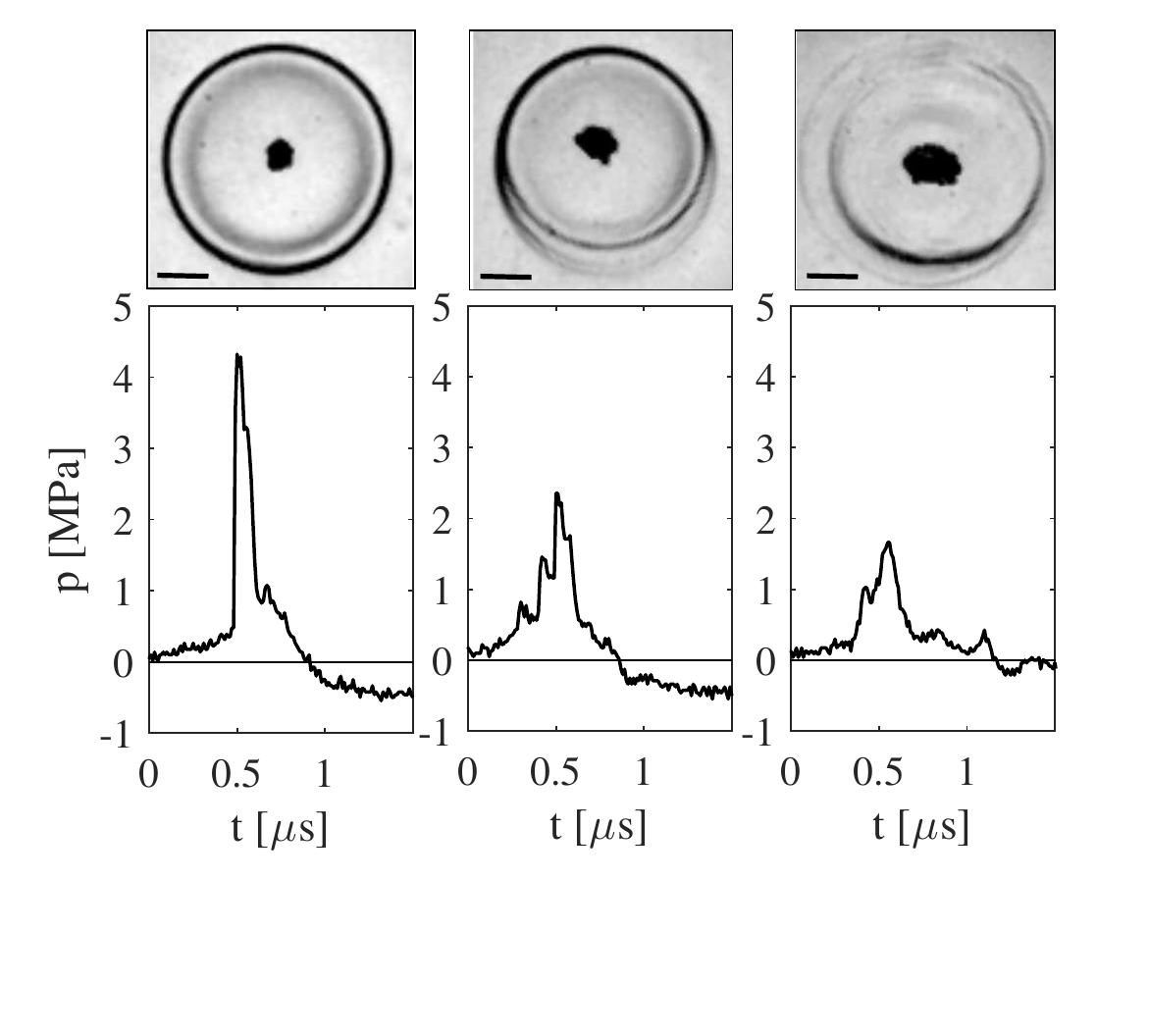}
	\put (14,42) {(a)}
	\put (41.5,42) {(b)}
	\put (69.5,42) {(c)}
\end{overpic}
\end{center}
\caption{Examples of hydrophone pressure signals of shock waves measured at the collapse of bubbles deformed by gravity at (a) $\zeta<10^{-3}$, (b) $\zeta = 3.8\times10^{-3}$ and (c) $\zeta = 10^{-2}$. The corresponding shadowgraph images with an exposure of 50~ns are shown on top. The black bars show the 1~mm scale.}
\label{fig:hydros}
\end{figure}

\begin{figure}
\begin{center}
\includegraphics[width=.7\textwidth, trim=0cm 2.3cm 0cm 0cm, clip]{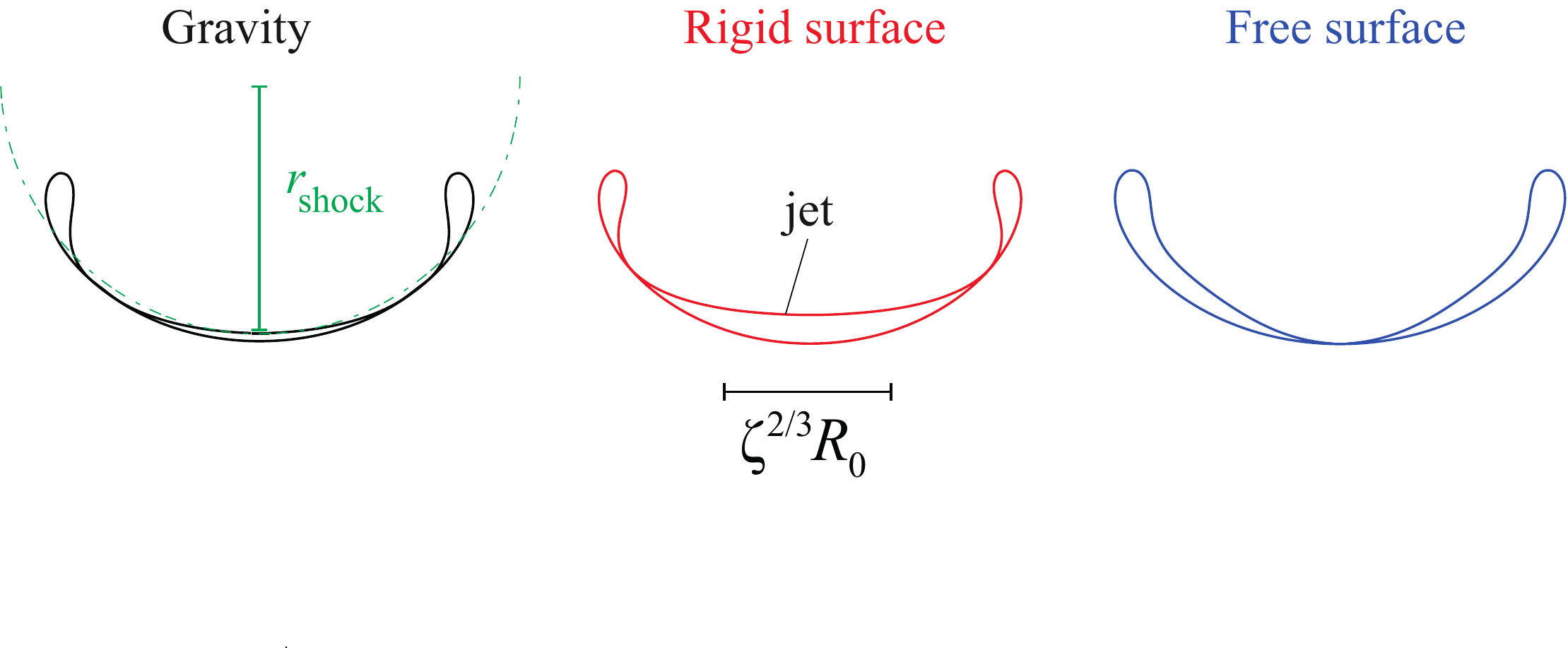}
\end{center}
\caption{Bubble shapes at jet impact for bubbles deformed by a uniform pressure gradient, a near rigid and a near free surface, predicted by potential flow theory~\cite{Supponen2016} for $\zeta=10^{-2}$. $\boldsymbol{\zeta}$ is directed downward.}
\label{fig:shapes}
\end{figure}

We now investigate shock waves from non-spherically collapsing bubbles at a more general level with the aim of developing an empirical model to predict their strengths.
For this purpose, we look at shock waves from bubbles deformed by different sources, in particular by the gravity-induced uniform pressure gradient.
Examples of measured shock waves from bubbles deformed by gravity are shown in Fig.~\ref{fig:hydros}.
A spherical collapse (Fig.~\ref{fig:hydros}a) produces a single shock, as observed previously in Section~\ref{s:spherical}.
Non-spherical collapses (Fig.~\ref{fig:hydros}b,c) generate multiple shocks, and the associated peak pressures clearly decrease with increasing bubble deformation, similarly to bubbles deformed by a free surface.
However, the characteristic shape of bubbles collapsing in uniform pressure gradients is such that the radii of curvature of the jet tip and the opposite bubble wall at their impact are very similar for a wide range of $\zeta$ according to potential flow theory~\cite{Supponen2016}, as illustrated in Figure~\ref{fig:shapes} for $\zeta=10^{-2}$.
As a consequence, the volumes of the `tip bubble' and the toroidal bubble remain relatively small and the associated shocks are barely distinguishable.
We therefore analyze the collapse shock as one event, expected to be dominated by the jet impact (as suggested by Figure~\ref{fig:z_E_all} for bubbles near a free surface at $\zeta<10^{-2}$), without resolving its sub-structure in the following analyses.

\begin{figure}
\begin{center}
\includegraphics[width=0.6\textwidth, trim=0cm 0cm 0cm 0cm, clip]{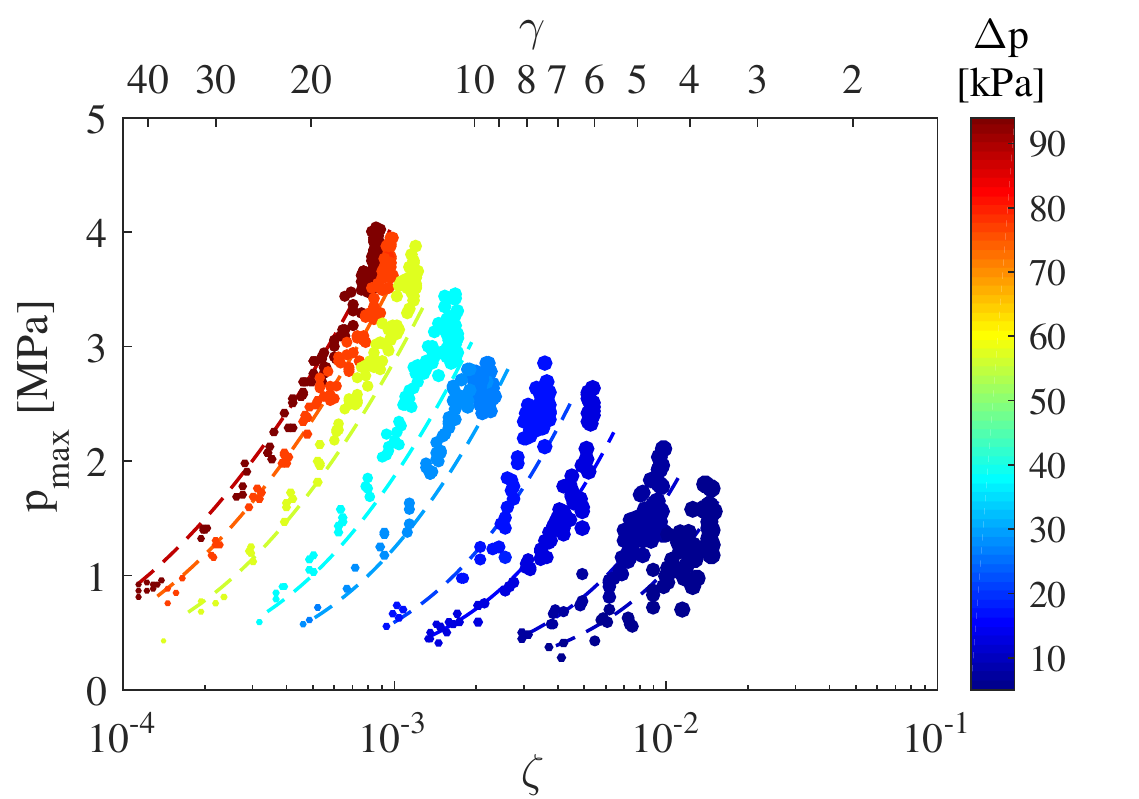}
\end{center}
\caption{Measured shock peak pressures as a function of $\zeta$ (and $\gamma$, top axis) for bubbles deformed by gravity. The dashed lines represent the model in Eq.~(\ref{eq:pmax}). The colors indicate different driving pressures $\Delta p$. The symbol sizes portray the different maximum bubble radii.}
\label{fig:z_hammer_g}
\end{figure}

We first consider the variation of the peak pressures $p_{\rm max}$ measured by the hydrophone as a function of $\zeta$. 
Figure~\ref{fig:z_hammer_g} shows this function for bubbles deformed by the gravity-induced pressure gradient (varied parameters: $R_{0}=1.5$--10~mm, $\Delta p=6$--98~kPa, at normal gravity). 
Clearly, the relation between $p_{\rm max}$ and $\zeta$ depends on $\Delta p$. 
We can build a model for the relationship between $p_{\rm max}$, $\Delta p$ and $\zeta$, based on the simplistic assumptions of scale-free micro-jets and shocks resulting from a water hammer pressure caused by the jet impact~\cite{Field1991,Johnsen2009}:
%
\begin{equation}
p_{h} = \frac{1}{2}\rho c U_{\rm jet} = 0.45\left(\rho c^{2}\Delta p\right)^{1/2} \zeta^{-1}
\label{eq:p_hammer}
\end{equation}
where $U_{\rm jet}$ is the micro-jet speed at its impact on the opposite bubble wall.
The scaling model for the micro-jet speed,  $U_{\rm jet} = 0.9 \left(\Delta p/\rho\right)^{1/2} \zeta^{-1}$, has previously been established by combining numerical simulations and analytical arguments with experimental observations, and is a valid approximation for jets driven by gravity and near surfaces at $\zeta < 0.1$~\cite{Supponen2016}.
We can therefore expect also the resulting hammer pressures to be similar for these different sources of bubble deformation and to decrease with $\zeta$ for a given $\Delta p$ (with constant $\rho$ and $c$).  
The scaling factor in Eq.~(\ref{eq:p_hammer}) could be different if the jet impact is not the dominant shock mechanism, but this is irrelevant in the following derivation because of the free parameter $\alpha$ discussed hereafter.
%

The equivalent observational proxy for $p_{h}$ is expressed as
%
\begin{equation}
p_{h} = p_{\rm max} \left(\frac{d}{r_{\rm shock}}\right)^{\beta} = \alpha p_{\rm max}\left(\frac{d}{R_{0}}\right)^{\beta} \zeta^{-2\beta/3} 
\label{eq:exp_p}
\end{equation}
where $p_{\rm max}$ is the peak pressure measured by the hydrophone, $d$ is the distance between the bubble center and the hydrophone sensor, $r_{\rm shock}$ is the shock emitting radius, assumed to scale as the radius of the jet tip (see schematic in Fig.~\ref{fig:shapes}) and thereby as the bubble's characteristic length at jet impact $s \propto \zeta^{2/3}R_ {0}$ as predicted by potential flow theory for $\zeta<<1$~\cite{Supponen2016}, and $\alpha$ and $\beta$ are free parameters.
$\alpha$ represents the unknown scaling of $r_{\rm shock}\propto \zeta^{2/3}R_ {0}$.
$\beta$ would equal 1 for negligible shock dissipation and spreading of the shock width, yet in reality non-linearities are present and result in a higher exponent, typically about 2 in the near field and $\sim1.1$ in the far field of the emission center~\cite{Schoeffmann1988,Doukas1991,Vogel1996,Pecha2000}.
Equating Eq.~(\ref{eq:p_hammer}) and~(\ref{eq:exp_p}) gives
%
\begin{equation}
p_{\rm max} = \frac{0.45}{\alpha} \left(\rho c^{2}\Delta p\right)^{1/2}\left(\frac{R_{0}}{d}\right)^{\beta}\zeta^{2\beta/3-1}.
\label{eq:pmax}
\end{equation}
We fit $\alpha$ and $\beta$ simultaneously to a sample of 931 bubbles deformed by gravity to minimize the $\chi^{2}$ deviation between the left and right hand sides of Eq.~(\ref{eq:pmax})~\footnote{A fit with the exponent of $\rho c^{2} \Delta p$ as a free parameter was also performed, which consistently gave $0.506 \pm 0.006$, which is why we kept this exponent as 1/2.}.
The resulting fitted parameters are $\alpha = 0.277 \pm 0.006$ and $\beta=1.249 \pm 0.003$, and the corresponding determination coefficient is $R^{2}=0.93$.
$\beta$ lies between 1 and 2 as expected.
In the case of bubbles deformed by gravity, there is a unique relation between $R_{0}$, $\Delta p$ and $\zeta$ as shown by Eq.~(\ref{eq:zeta}).
Substituting $R_{0}$ from this relation into Eq.~(\ref{eq:pmax}) makes $p_{\rm max}$ a function of only $\Delta p$ and $\zeta$.
These relations are plotted as dashed lines in Fig.~\ref{fig:z_hammer_g} and show excellent agreement with the measurements.

\begin{figure}
\begin{center}
\includegraphics[width=0.6\textwidth, trim=0cm 0cm 0cm 0cm, clip]{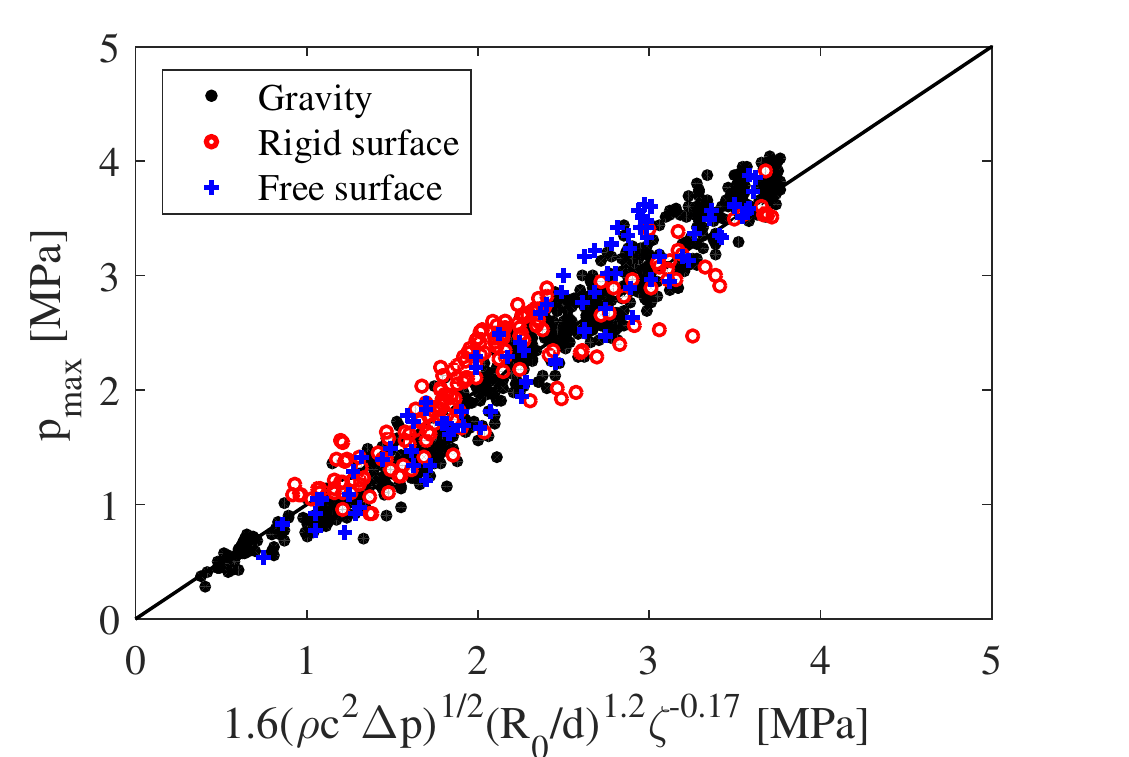}
\end{center}
\caption{Measured shock wave peak pressures as a function of the model given in Eq.~(\ref{eq:pmax}) for bubbles deformed by gravity, a rigid and a free surface.}
\label{fig:z_hammer_all}
\end{figure}
%

The lines in Fig.~\ref{fig:z_hammer_g} can be collapsed to a single relationship by plotting the measured peak pressures $p_{\rm max}$ directly against the model in Eq.~(\ref{eq:pmax}), which is shown in Fig.~\ref{fig:z_hammer_all}.
We now also apply this simple model to predict the shock pressures of non-spherical bubbles with different sources of deformation (free and rigid surface), where the unique relationship between $R_{0}$, $\Delta p$ and $\zeta$ no longer holds because of the additional dependence on the distance $h$ to the surface, as shown by Eq.~(\ref{eq:zeta}).
These data also coincide with the model, as seen in Fig.~\ref{fig:z_hammer_all}, confirming that the hammer pressure model can be used to estimate shock pressures produced by a non-spherical bubble collapse.
The pressures $p_{h}$ at the source, estimated using Eq.~(\ref{eq:p_hammer}) and~(\ref{eq:exp_p}), range from 100~MPa to 10~GPa at $\zeta > 10^{-3}$.

\begin{figure}
\begin{center}
\includegraphics[width=\textwidth]{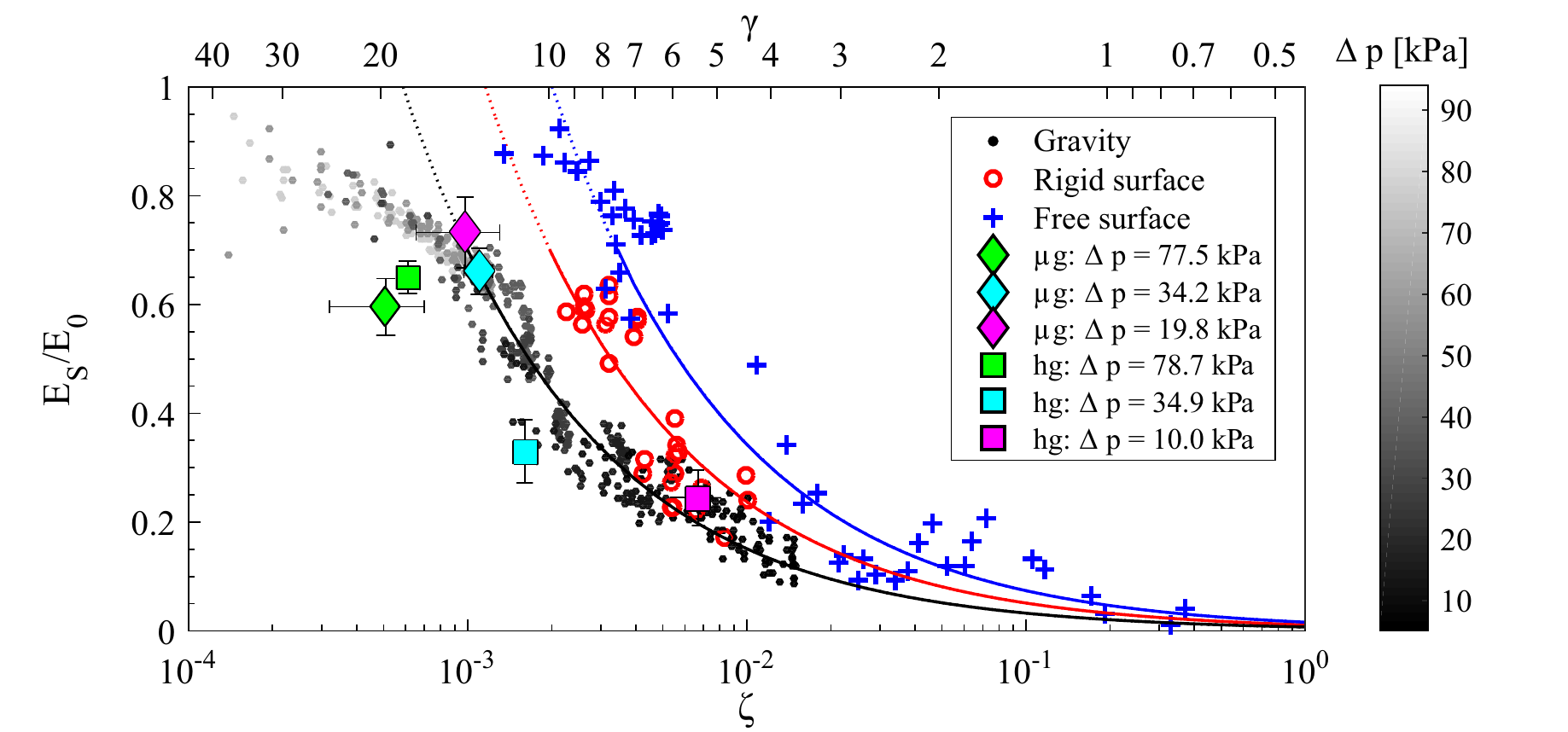}
\end{center}
\caption{The normalized total collapse shock wave energy for bubbles deformed by gravity, a near rigid and a near free surface as a function of $\zeta$ (and $\gamma$, top axis). Averaged shock energies measured in micro-gravity $\mu$g ($0\pm0.02$~$g$) and hyper-gravity hg ($1.66\pm0.093$~$g$) at three different $\Delta p$ are also displayed. The gray scale indicates different driving pressures $\Delta p$ for bubbles deformed by gravity. The models in solid lines show the fits $0.0073\zeta^{-2/3}$, $0.011\zeta^{-2/3}$ and $0.016\zeta^{-2/3}$ for bubbles deformed by gravity, rigid surface and free surface, respectively. The mean error of $E_{S}/E_{0}$ is 0.04.}
\label{fig:z_E_general}
\end{figure}

Figure~\ref{fig:z_E_general} displays the normalized collapse shock wave energy for bubbles deformed by gravity, a nearby rigid and a free surface as a function of $\zeta$.
All the measured shock energies generally decrease with increasing $\zeta$ independently of $R_{0}$ and $\Delta p$.
For gravity-deformed bubbles, most of the decrease happens in the range $10^{-3}<\zeta<10^{-2}$, reaching values down to about 10\% of initial bubble energy $E_{0}$ at $\zeta \sim 10^{-2}$.
These values differ significantly from bubbles deformed by a rigid and a free surface that respectively have shock energies as high as 30\% and 40\% of the initial bubble energy $E_{0}$ at $\zeta \sim 10^{-2}$ ($\gamma \sim 4.4$).
Shocks from bubbles deformed by a near rigid and a free surface experience a decrease in energy with $\zeta$ that is similar to the gravity-deformed cases, but which occurs at a higher $\zeta$.

It should be noted that the expression of $\zeta$ for gravity-induced bubble deformations (Eq.~(\ref{eq:zeta})) includes $\Delta p$, making $\Delta p$ correlate with $\zeta$ in our data obtained on-ground (see gray scale in Fig.~\ref{fig:z_E_general}).
However, the data in micro-gravity ($0\pm0.02$~$g$), which were obtained aboard ESA parabolic flights, confirm that the bubble deformation is the main cause of the observed shock energy variations, rather than $\Delta p$.
For example, bubbles collapsing at $\Delta p\approx20$~kPa in our experiment on-ground emit low-energy shocks ($E_{S}/E_{0}<30\%$), yet in micro-gravity at the same driving pressure $E_{S}/E_{0}>75\%$~\footnote{The presence of the closest surface to the bubble, i.e.~the parabolic mirror, is accounted for when determining $\zeta$ for bubbles collapsing in micro-gravity.}.
Some data for bubbles collapsing at higher gravity levels ($1.66\pm0.093$~$g$) are also displayed in Figure~\ref{fig:z_E_general}, showing reasonable agreement with the general shock energy trend with $\zeta$.

Since the measured peak pressures for deformed bubbles are well approximated with the hammer pressure model, we aim at estimating their shock energies using the same approach.
We recall that the shock energy $E_{S} = (4\pi d^{2}\rho^{-1}c^{-1})\int p^{2}\text{d}t$ from ref.~\cite{Vogel1988}, as for Eq.~(\ref{eq:ES}).
If the pressure profile in time is represented with a hammer pressure $p_{h}$ being applied for a time $\Delta t=\Delta d c^{-1}$, where $\Delta d$ denotes the thickness of the shock, the energy reads $E_{S} = (4\pi d^{2}\rho^{-1}c^{-1})p_{h}^{2} \Delta t$.
The shock wave energy is therefore alternatively expressed as
%
\begin{equation}
E_{S} = \frac{\Delta V p_{h}^{2}}{\rho c^{2}},
\label{eq:ES2}
\end{equation}
where $\Delta V=4\pi d^{2}\Delta d$ is the volume of the compressed liquid.
%
As mentioned before, the characteristic length of the bubble at the jet impact scales as $s/R_{0} \propto \zeta^{2/3}$. 
As the surface area of contact of the jet onto the opposite bubble wall is two-dimensional and the compressed liquid volume is assumed to be proportional to that area, we have $\Delta V/R_{0}^{3} \propto s^{2}/R_{0}^{2} \propto \zeta^{4/3}$.
With this model plugged into Eq.~(\ref{eq:ES2}) and $p_{h}$ substituted for Eq.~(\ref{eq:p_hammer}), we obtain
\begin{equation}
\frac{E_{S}}{E_{0}} \propto \frac{\Delta V}{R_{0}^{3}\zeta^{2}}  \propto \zeta^{-2/3}.
\label{eq:eps}
\end{equation}
%
The missing scaling factor for Eq.~(\ref{eq:eps}) comes from the unknown size of the compressed liquid region.
An analytical evaluation of this unknown is difficult and would have to account for the non-uniform liquid compression by the curved jet tip.
The scaling factor is expected to vary for the distinct sources of deformations, since the jet shapes are different for each case and leave gas/vapor pockets of dissimilar sizes between the jet and the opposite bubble wall, as illustrated in Fig.~\ref{fig:shapes} for $\zeta = 10^{-2}$.
These vapor pockets are rather large for bubbles collapsing near a rigid or a free surface, while gravity-induced jets hit the opposite bubble wall in a highly uniform way, thereby resulting in the smallest scaling factor.
When minimizing the $\chi^{2}$ deviation between the measurements $E_{S}/E_{0}$ for bubbles deformed by gravity at $\zeta > 10^{-3}$ and a model in the form $f = a\zeta^{b}$ with free parameters $a$ and $b$, we find $a = 0.0078$ and $b = -0.66$. When imposing $b=-2/3$ to conform with Eq.~(\ref{eq:eps}), the best fit for $a$ is 0.0073.
The corresponding fitted scaling factor for the rigid and free surface are $a=0.011$ and 0.016, respectively.
Equation~(\ref{eq:eps}) with these fitted scaling factors is plotted as solid lines for bubbles deformed by gravity, free surface and rigid surface in Fig.~\ref{fig:z_E_general}, and agrees reasonably well with the experimental data.

\section{Discussion}
\label{s:discussion}

There are several limitations in the presented shock models worth addressing.
The micro-jet is expected to reach the speed of sound for a bubble collapsing at $\zeta \lesssim 0.9(\Delta p/\rho)^{1/2}c^{-1}$ ($\zeta \lesssim 0.006 $ at $\Delta p = 98$~kPa), below which the model in Eq.~(\ref{eq:p_hammer}) may no longer be able to estimate the jet hammer pressures.
Furthermore, our model neglects the gas inside the bubble. 
Compressed and heated gases within highly spherically collapsing bubbles can potentially slow down and destroy the jet and/or delay or prevent its formation. 
These effects naturally decrease with increasing $\zeta$, since at higher $\zeta$ the jet forms earlier in the bubble evolution, when the gases are less compressed. 
We estimate the bubble gas to seriously hamper the jet for $\zeta<10^{-3}$, where no observable jets are formed in the bubble rebound in our current setup~\cite{Supponen2016}.
This is the likely explanation for the sudden curvature change in the shock energy trend for bubbles deformed by gravity at $\zeta \sim 10^{-3}$, as seen in Fig.~\ref{fig:z_E_general}.
Below this approximate threshold (at which $p_{h}\sim7$~GPa for bubbles collapsing here at atmospheric pressure), the shock pressures predicted by the model are overestimated.
This threshold value is consistent with previous findings for a spherical collapse at atmospheric pressure, both in our setup (Section~\ref{s:spherical}) and in the literature~\cite{Pecha2000,Akhatov2001,Holzfuss1998}.

The shock energies of bubbles collapsing near a rigid surface show important differences when compared with the measurements performed by Vogel and Lauterborn~\cite{Vogel1988}.
Although they observed, similarly to us with bubbles near a free surface, a clear minimum in shock energies at $\gamma = 1$, they also measured shocks beyond $\gamma \sim 3$ to have the same energies as those emitted in a spherical collapse, while at $\gamma=3$ we measure barely $20\%$ of a typical shock energy from a spherical collapse.
It suggests that the experimental conditions play an important role on the collapse shock wave characteristics, including the initial bubble sphericity, which highly differs for parabolic mirror and lens-based laser focusing methods.
Indeed, in Vogel's study the stand-off was varied only up to $\gamma \sim 3$, beyond which a spherical collapse was assumed, while we still find important shock energy variations between $\gamma \sim 5$ and 10.

\section{Conclusion}
\label{s:conclusion}

We have presented detailed observations of shock wave emissions from the collapse of bubbles with various levels of deformation, quantified by the anisotropy parameter $\zeta$, using simultaneous time-resolved shadowgraphy and needle hydrophone pressure measurements.
A gradual pressure rise in the liquid near the bubble wall was observed in the last collapse stage of nearly spherically collapsing bubbles, in agreement with the century-old predictions of Lord Rayleigh.
Non-spherical bubble collapses produced multiple shock waves associated with different processes such as the jet impact and the individual collapses of the various separated parts of the bubble. 
When quantifying these distinct shocks for bubbles collapsing near a free surface, the jet impact shock was found to dominate up to $\zeta \sim 2\times10^{-2}$, the bubble tip collapse in the range $2\times10^{-2} < \zeta < 0.15$ and the torus collapse at $\zeta > 0.15$.
The timings of the individual events, normalized with the bubble collapse time, were also found to vary with $\zeta$.

Models predicting the shock peak pressure and energy were proposed based on the assumption that the shock wave is generated by a jet impact hammer pressure.
The pressure model showed excellent agreement with the observed data in the range $10^{-3}<\zeta<10^{-2}$ for all three sources of bubble deformation used here (gravity, rigid and free surface), and the energy model captured the approximative trend of the measured energies.
The total collapse shock wave energy, normalized to the total bubble energy, generally decreased with increasing $\zeta$.
However, we found differences between the shock energies from bubbles deformed by different sources, which likely result from the small variations in the jet shapes at their impact onto the opposite bubble wall.
Interestingly, these differences do not seem to affect the shock peak pressures, which could be due to the jet speed at the moment of impact - which the hammer pressure is proportional to - being nearly identical for the three sources of bubble deformation at this range of $\zeta$.

\acknowledgements{We acknowledge the support of the Swiss National Science Foundation (Grant no. 513234), the University of Western Australia Research Collaboration Award obtained by co-authors DO and MF, the European Space Agency for the 60th and 62nd parabolic flight campaigns and Prof.\ Ullrich for the 1st Swiss parabolic flight.}

\bibliography{mybib}

\end{document}